\documentclass[3p]{elsarticle}

\usepackage{paralist}
\usepackage{xcolor}
\usepackage{times}
\usepackage{bmpsize}
\usepackage{latexsym}
\usepackage{hyperref}
\usepackage{paralist}
\usepackage{ifthen}
\usepackage{xspace}
\usepackage{amsmath,amssymb}
\usepackage[ruled,linesnumbered]{algorithm2e}
\usepackage{caption}
\usepackage{subcaption}
\usepackage{float}

\usepackage{lscape}

\definecolor{vuborange}{rgb}{1.0,0.40,0.0}

\newcommand\m[1]{\ensuremath{#1}\xspace}

\newcommand\leqp{\m{\leq_p}}
\newcommand\geqp{\m{\geq_p}}
\newcommand\entails{\m{\models}}
\newcommand\limplies{\m{\Rightarrow}}
\newcommand\ourtool{\textsc{ZebraTutor}\xspace}
\newcommand\idp{\textsc{IDP}\xspace}
\usepackage{amsthm}
\usepackage{thmtools} 

\newtheorem{thm}{Theorem}
\newtheorem{definition}[thm]{Definition}

\newcommand\ddd{\m{\overline{d}}}

\newcommand{\myparagraph}[1]{\subsection{#1}}

\newcommand\allconstraints{\m{T}}

\journal{Journal of Artificial Intelligence}

\begin{document}

\begin{frontmatter}

\title{A framework for step-wise explaining how to solve constraint satisfaction problems}

\author[mymainaddress]{Bart Bogaerts, Emilio Gamba, Tias Guns}
\address{Vrije Universiteit Brussel, Pleinlaan 2, 1050 Brussel, Belgium}
\ead{\{firstname.lastname\}@vub.be}
\date{}

\begin{abstract}
We explore the problem of step-wise explaining how to solve constraint satisfaction problems, with a use case on logic grid puzzles.
More specifically, we study the problem of explaining the inference steps that one can take during propagation, in a way that is easy to interpret for a person.
Thereby, we aim to give the constraint solver explainable agency, which can help in building trust in the solver by being able to understand and even learn from the explanations.
The main challenge is that of finding a sequence of \textit{simple} explanations, where each explanation should aim to be as cognitively easy as possible for a human to verify and understand. 
This contrasts with the arbitrary combination of facts and constraints that the solver may use when propagating.
We propose the use of a cost function to quantify how simple an individual explanation of an inference step is, and identify the explanation-production problem of finding the best sequence of explanations of a CSP. 
Our approach is agnostic of the underlying constraint propagation mechanisms, and can provide explanations even for inference steps resulting from combinations of constraints. 
In case multiple constraints are involved, we also develop a mechanism that allows to break the most difficult steps up and thus gives the user the ability to \emph{zoom in} on specific parts of the explanation. 
Our proposed algorithm iteratively constructs the explanation sequence by using an optimistic estimate of the cost function to guide the search for the best explanation at each step.
Our experiments on logic grid puzzles show the feasibility of the approach in terms of the quality of the individual explanations and the resulting explanation sequences obtained.
\end{abstract}


\end{frontmatter}


\section{Introduction}\label{sec:intro}

In the last few years, as AI systems employ more advanced reasoning mechanisms and computation power, it becomes increasingly difficult to understand why certain decisions are made. 
Explainable AI (XAI) research aims to fulfil the need for trustworthy AI systems to understand why the system made a decision for verifying the correctness of the system, as well as to control for biased or systematically unfair decisions.

Explainable AI is often studied in the context of (black box) prediction systems such as neural networks, where the goal is to provide insight into which part(s) of the input is important in the \textit{learned} model. 
These insights (or local approximations thereof) can justify why certain predictions are made. 
In that setting, a range of techniques have been developed, ranging from local explanations of predictions at the \textit{feature level}~\cite{ribeiro2016should,lundberg2017unified} to \textbf{visual explanations} with \textit{saliency maps} \cite{selvaraju2017grad}. 
Adadi et al.~\cite{Adadi_2018}, Guidotti et al.~\cite{guidotti2018survey} and more recently Arrieta et al.~\cite{Barredo_Arrieta_2020}, survey the latest trends and  major research directions in this area.
In contrast, in Constraint Satisfaction Problems (CSP) \cite{fai/Rossi06}, the problem specification is an explicit \textit{model-based representation} of the problem, hence creating the opportunity to explain the inference steps directly in terms of this representation.

Explanations have been investigated in constraint solving before, most notably for explaining overconstrained, and hence unsatisfiable, problems to a user~\cite{junker2001quickxplain}.
Our case is more general in that it also works for satisfiable problems.
At the solving level, in lazy clause generation solvers, explanations of a constraint are studied in the form of an implication of low-level Boolean literals that encode the result of a propagation step of an individual constraint~\cite{feydy2009lazy}. 
Also, no-goods (learned clauses) in conflict-driven clause learning SAT solvers can be seen as explanations of failure during search~\cite{marques2009conflict}. 
These are not meant to be human-interpretable but rather to propagate effectively.

We aim to explain the process of propagation in a constraint solver, independent of the consistency level of the propagation and without augmenting the propagators with explanation capabilities.
For problems that can --- given a strong enough propagation mechanism --- be solved without search, e.g. problems such as logic grid puzzles with a unique solution, this means explaining the entire problem solving process. 
For problems involving search, this means explaining the inference steps  in one search node. 
It deserves to be stressed that we are not interested in the computational cost of performing an expensive form of propagation, but in explaining all consequences of a given assignment to the user in a way that is as understandable as possible. 

More specifically, we aim to develop an explanation-producing system that is complete and interpretable. 
By \emph{complete} we mean that it finds a \emph{sequence} of small reasoning steps that, starting from the given problem specification and a partial solution, derives all consequences. 
Gilpin et al.~\cite{DBLP:conf/dsaa/GilpinBYBSK18} define \emph{interpretable} explanations as ``descriptions that are simple enough for a person to understand, using a vocabulary that is meaningful to the user''. 
Our guiding principle is that of simplicity, where smaller and simpler explanations are better. 

An open issue when presenting a sequence of explanations is the level of abstraction used.
Starting from a sequence of as-simple-as-possible explanations as we do, there are two ways to change the level of abstraction.
The first direction is that one could \emph{group} multiple single steps into one larger step that can then be expanded on the fly. 
The second direction is to provide a \emph{more detailed} explanation of harder steps which can not be broken into simpler steps with the standard mechanism.
While the first direction can be useful when explanations get very long, from a theoretical perspective, it is less interesting. 
The second direction on the other hand is interesting for several reasons. First of all, we noticed that some of the generated steps are still too complicated to be understood easily and thus really require being explained in more detailed. Secondly, breaking them up in smaller steps is an interesting challenge. Indeed, since we start from the idea that steps should be as simple as possible, it should not be possible to break them up further. 
To still break them up further, we took inspiration from methods people use when solving puzzles, and mathematicians use when proving theorems. 
That is, for explaining a single step in more detail we work using reasoning by contradiction: starting from the negation of a consequence, we explain how a contradiction is obtained.  In essence, we assume the negation of the derived fact, and can then reuse the principle of explanation steps to construct a sequence that leads to an inconsistency. 
This novel approach allows us to provide a mechanism for \emph{zooming in} on the most difficult explanation step.

\begin{figure}[h]
\centering
\includegraphics[width=\textwidth]{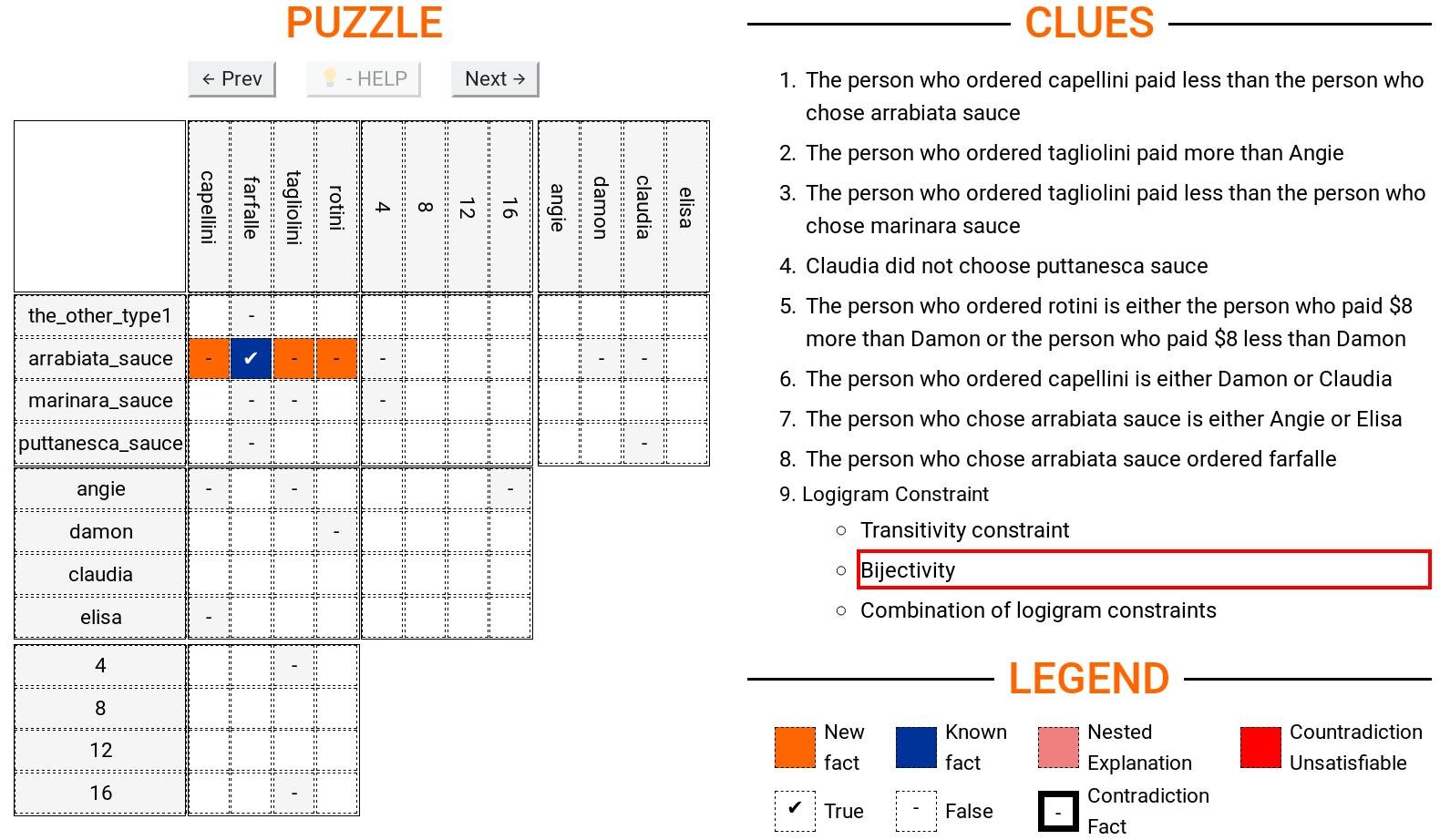}
\caption{Demonstration of explanation visualisation.}
\label{fig:zebrascreen}
\end{figure}

In practice, we choose to present the constraints in natural language, which is an obvious choice for logic grid puzzles as they are given in the form of natural language \textit{clues}. 
We represent the previously and newly derived facts visually, as can be seen in the grid in Figure~\ref{fig:zebrascreen}. In this figure, the implicit ``Bijectivity'' axiom present in each logic grid puzzle is used to derive the following: since Arrabiata sauce was eaten with Farfalle, it was not eaten with any of the other pasta types.

Our work and more specifically the use case of logic grid puzzles is motivated by the ``Holy Grail Challenge''\footnote{\url{https://freuder.wordpress.com/pthg-19-the- third-workshop-on-progress-towards-the-holy-grail/}} which had as objective to provide \textit{automated} processing of logic grid puzzles, ranging from natural language processing, to solving, and explaining.
While our integrated system, \ourtool, has the capability of solving logic grid puzzle starting from the natural language clues (see Section \ref{sec:holistic}), the focus in this paper is on the explanation-producing part of the system.

The explanation-generating techniques we develop can be applied in a multitude of use cases. 
For instance, it can explain the entire sequence of reasoning, such that a user can better understand it, or in case of an unexpected outcome can debug the reasoning system or the set of constraints that specify the problem. 
As our approach starts from an arbitrary set of facts, it can also be used as a virtual assistant when a user is stuck in solving a problem.
The system will explain the simplest possible next step, or in an interactive setting, the system can explain how it would complete a partial solution of a user. Such explanations can be useful in the context of  \emph{interactive configuration} \cite{felfernig2014knowledge} where a domain expert solves a problem (e.g., a complicated scheduling or rostering problem) but is assisted by an intelligent system. 
The system in that case typically does not have the all the knowledge required to solve the problem since certain things are hard to formalize, especially when personal matters are involved. In such case, the system cannot find the optimal solution automatically, but it can help the expert for instance by propagating information that follows from its knowledge base. In case something undesired is propagated, the user might need an explanation about \emph{why} this follows; this is where our methods can be plugged in.
Finally, our measures of simplicity of reasoning steps can be used to estimate the difficulty of solving a problem for a human, e.g. for gradual training of experts.

Summarized, our main contributions are the following:
\begin{itemize}
	\item We formalize the problem of step-wise explaining the propagation of a constraint solver through a sequence of small inference steps;
	\item We propose an algorithm that is agnostic to the propagators and the consistency level used, and that can provide explanations for inference steps involving arbitrary combinations of constraints;
	\item Given a cost function quantifying human interpretability, our method uses an optimistic estimate of this function to guide the search to low-cost explanations, thereby making use of Minimal Unsatisfiable Subset extraction;
	\item We introduce nested explanations to provide additional explanations of complex inference steps using reasoning by contradiction;
	\item We experimentally demonstrate the quality and feasibility of the approach in the domain of logic grid puzzles.
\end{itemize}

This paper is structured as follows. In Section~\ref{sec:related-work}, we discuss related work. Section \ref{sec:background}, explains the rules of logic grid puzzles and presents background information. 
Sections \ref{sec:problem-definition} and \ref{sec:nested-explanation}, formalize the theory of the explanation-production problem on two abstraction levels, while Section \ref{sec:expl-gen-prod} describes the algorithms needed to solve the explanation-production problem.
In Section \ref{sec:zebra}, we motivate our design decisions using observations from the \ourtool use case. 
Section \ref{sec:holistic} describes the entire information pipeline of the \ourtool integrated system. 
In Section \ref{sec:experiments}, we experimentally study the feasibility of our approach. 
Finally, we conclude the paper in Section \ref{sec:conclusion}.

\paragraph{Publication history} This paper is an extension of previous papers presented at workshops and conferences \cite{claesuser,DBLP:conf/bnaic/ClaesBCGG19,ecai/BogaertsGCG20}. The current paper extends the previous papers with more detailed examples, additional experiments, as well as the formalization of what we call \emph{nested explanation sequences}.

\section{Related work}\label{sec:related-work}
This research fits within the general topic of Explainable Agency~\cite{langley2017explainable}, whereby in order for people to trust autonomous agents, the latter must be able to \textit{explain their decisions} and the \textit{reasoning} that produced their choices.

Explanation of Constraint Satisfaction Problems (CSP) has been studied most in the context of over-constrained problems.
The goal then is to find a small conflicting subset.
The QuickXplain method \cite{junker2001quickxplain} for example uses a dichotomic approach that recursively partitions the constraints to find a minimal conflict set. Many other papers consider the same goal and search for explanations of over-constrainedness~\cite{leo2017debugging,zeighami2018towards}.
A minimal set of conflicting constraints is often called a \emph{minimal unsatisfiable subset} (MUS) or \emph{minimal unsatisfiable core} \cite{marques2010minimal}. Despite the fact that we do not (specifically) aim to explain overconstrained problems, our algorithms will internally also make use of MUS extraction methods.

While explainability of constraint optimisation has received little attention so far, in the related field of \textit{planning}, there is the emerging subfield of \textit{eXplainable AI planning} (XAIP)~\cite{fox2017explainable}, which is concerned with building planning systems that can explain their own behaviour. This includes answering queries such as ``why did the system (not) make a certain decision?'', ``why is this the best decision?'', etc. In contrast to explainable machine learning research~\cite{guidotti2018survey}, in explainable planning one can make use of the explicit \textit{model-based representation} over which the reasoning happens. Likewise, we will make use of the constraint specification available to constraint solvers, more specifically typed first-order logic~\cite{atcl/Wittocx13}.

\begin{figure}[ht]
    \centering
    \includegraphics[width=\textwidth]{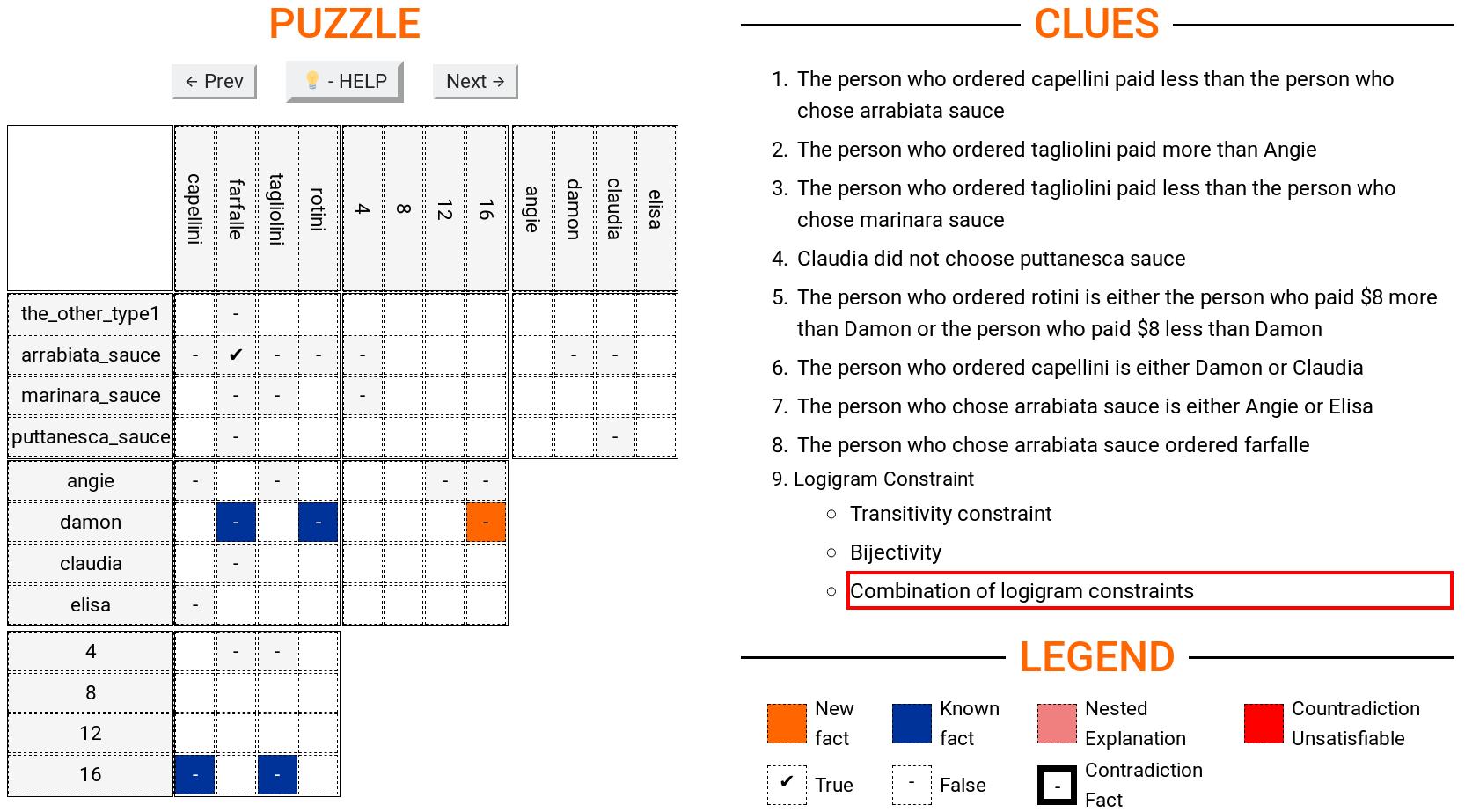}
    \caption{Demonstration of explanation visualisation.}
    \label{fig:zebrascreen:path}
\end{figure}

Our work was inspired by the Holy Grail Challenge \cite{freuder2018progress} at the 2019 Constraint Programming conference (CP), which in turn has its roots in earlier work of E.~Freuder on inference-based explanations \cite{sqalli1996inference}.
In the latter, the authors investigate logic grid puzzles and develop a number of problem-specific inference rules that allow solving (most, but not all) such puzzles without search.
These inference rules are equipped with explanation templates such that each propagation event of an inference rule also has a templated explanation, and hence an explanation of the solution process is obtained.
We point out that the more complex inference rules (NCC and GNCC) are in fact inference rules over hard-coded combinations of (in)equality constraints.
In contrast, our proposed method works for any type of constraint and any combination of constraints, and automatically infers a minimal set of facts and constraints that explain an inference step, without using any problem-specific knowledge.
This powerful combination of constraints is able to automatically detect interesting consistency patterns that needed to be hand-coded in the Freuder's seminal work, but also in the solutions submitted by other
contestants in the challenge \cite{escamocher2019solving}.
Figure \ref{fig:zebrascreen:path} shows an example of a non-trivial explanation that our approach automatically generated; a combination of a so-called bijectivity and transitivity constraints, which was hard-coded as the special-purpose 'path consistency' pattern in earlier logic-grid specific work~\cite{sqalli1996inference}.

There is a rich literature on automated and interactive theorem proving, recently focussing on providing proofs that are understandable for humans \cite{Ganesalingam2017} and, e.g.,  on teaching humans -- using interaction with theorem provers -- how to craft mathematical proofs \cite{DBLP:conf/icml/YangD19}.
Our work fits into this line of research since our generated explanations can also be seen as proofs, but in the setting of finite-domain constraint solving.

Our approach also relates to the work of Belahcene et.\ al.~\cite{belahcene2017explaining} who --- in the context of decision-aiding --- aim to build incremental explanations using preference relations. In our case, this would correspond to preferring simple constraints to more complex combinations of constraints through a cost function.

\section{Background}\label{sec:background}\label{sec:prelims}
While our proposed method is applicable to constraint satisfaction problems in general, we use \textit{logic grid puzzles} as example domain, as it requires no expert knowledge to understand.
Our running example is a puzzle about people having dinner in a restaurant and ordering different types of pasta, which is the hardest logic grid puzzle we encountered; it was sent to us by someone who got stuck solving it and wondered whether it was correct in the first place.    
The entire puzzle can be seen in Figure \ref{fig:zebrascreen}; the full explanation sequence generated for it can be found online at \url{http://bartbog.github.io/zebra/pasta}.

In this section, we first present logic grid puzzles as a use case, and afterwards introduce \emph{typed first-order logic}, which we use as the language to express our constraint programs in. Here, it is important to stress that our definitions and algorithms work for any language with model-theoretic semantics, including typical constraint programming languages~\cite{rossi2006handbook}.

\subsection{Logic grid puzzles}
A logic grid puzzle (also known as ``Zebra puzzle'' or ``Einstein puzzle'') consists of natural language sentences (from hereon referred to as ``clues'') over a set of \emph{entities} occurring in those sentences. 
For instance, our running example in Figure~\ref{fig:zebrascreen} contains as second clue ``The person who chose arrabiata sauce is either Angie or Elisa'' and (among others) the entities ``arrabiata sauce'', ``Angie'' and ``Elisa''. 

The set of entities is sometimes left implicit if it can be derived from the clues, but often it is given in the form of a grid. 
Furthermore, in such a puzzle the set of entities is partitioned into equally-sized groups (corresponding to \emph{types}); in our example, ``person'' and ``sauce'' are two such types. 
The goal of the puzzle is to find relations between each two types such that
\begin{itemize}
	\item \textit{Each clue is respected}; 
	\item \textit{Each entity of one type is matched with exactly one entity of the second type}, e.g., each person chose exactly one sauce and each sauce is linked to one person (this type of constraint will be referred to as \emph{bijectivity}); and 
	\item \textit{The relations are logically linked}, e.g., if Angie chose arrabiata sauce and arrabiata sauce was paired with farfalle, then Angie must also have eaten farfalle (from now on called \emph{transitivity}). 
\end{itemize}

In Section \ref{sec:holistic}, we explain how we obtain a vocabulary and first-order theory in a mostly automated way from the clues using Natural Language Processing. 
The result is a vocabulary with types corresponding to the groups of entities in the clues, and the names and types of the binary relations to find (e.g \textit{chose(person, sauce)}, \textit{paired(sauce, pasta)}, \textit{eaten(person, pasta)});
as well as constraints (first-order sentences) corresponding to the clues, and the bijectivity and transitivity constraints. 

\subsection{Typed first-order logic}

Our constraint solving method is based on \emph{typed first-order logic}. 
Part of the input is a logical vocabulary consisting of a set of types, 
(typed) constant symbols, and (typed) relation symbols with associated type signature (i.e., each relation symbol is typed $T_1\times \dots \times T_n$ with $n$ types $T_i$).
\footnote{We here omit function symbols since they are not used in this paper.} 
For our running example, constant symbol \textit{Angie} of type \textit{person} is linked using relation \textit{chose(.,.)} with signature \textit{person $\times$ sauce} to constant symbol \textit{arrabiata sauce} of type \textit{sauce}.

A \emph{first-order theory} is a set of sentences (well-formed variable-free first-order formulas \cite{enderton} in which each quantified variable has an associated type), also referred to as constraints.
Since we work in a fixed and finite domain, the vocabulary, the interpretation of the types (the domains) and the constants are fixed.
This justifies the following definitions: 
\begin{definition}\label{def:partial-interpretation}
 A \emph{(partial) interpretation} is a finite set of literals, i.e., expressions of the form $P(\ddd)$ or $\lnot P(\ddd)$ where $P$ is a relation symbol typed $T_1\times\dots \times T_n$ and $\ddd$ is a tuple of domain elements where each $d_i$ is of type $T_i$. 
\end{definition}
 \begin{definition}\label{def:consistent}
 A partial interpretation is \emph{consistent} if it does not contain both an atom and its negation, it is called a \emph{full} interpretation if it either contains $P(\ddd)$ or $\lnot P(\ddd)$ for each well-typed atom $P(\ddd)$. 
\end{definition}

For instance, in the partial interpretation $I_1=\{chose(Angie,arrabiata),$ $\lnot chose(Elisa,arrabiata)\}$ it is known that $Angie$ had arrabiata sauce while Elisa did not. This partial interpretation does not specify whether or not Elisa ate Farfalle.

The syntax and semantics of first-order logic are defined as usual (see e.g.\ \cite{enderton}) by means of a satisfaction relation $I \models T$ between first-order theories $T$ and full interpretations $I$. If $I\models T$, we say that $I$ is a model (or solution) of $T$.

\begin{definition}
	A partial interpretation $I_1$ is \emph{more precise} than partial interpretation $I_2$ (notation $I_1\geqp I_2$) if $I_1\supseteq I_2$.
\end{definition}

Informally, one partial interpretation is more precise than another if it contains more information. For example, the partial interpretation $I_2 =\{chose(Angie,arrabiata)$, $\lnot chose(Elisa,arrabiata)$, $ \lnot chose(damon,arrabiata)\}$ is more precise than $I_1$ ($I_2 \geqp I_1$).

For practical purposes, since variable-free literals are also sentences, we will freely use a partial interpretation as (a part of) a theory in solver calls or in statements of the form $I\land T \models J$, meaning that everything in $J$ is a consequence of $I$ and $T$, or stated differently, that $J$ is less precise than any model $M$ of $T$ satisfying $M\geqp I$. 

In the context of first-order logic, the task of finite-domain constraint solving is better known as \emph{model expansion} \cite{MitchellTHM06}: given a logical theory $T$ (corresponding to the constraint specification) and a partial interpretation $I$ with a finite domain (corresponding to the initial domain of the variables), find a model $M$ more precise than $I$ (a partial solution that satisfies $T$).

If $P$ is a logic grid puzzle, we will use $T_P$ to denote a first-order theory consisting of:
\begin{itemize}
 \item One logical sentence for each clue in $P$;
 \item A logical representation of all possible bijection constraints;
 \item A logical representation of all possible transitivity constraints.
\end{itemize}

For instance, for our running example, some sentences in $T_P$ are: 
\begin{align*}
 &\lnot chose(claudia,puttanesca).\\
 & \forall s\in sauce : \exists p\in pasta : pairedwith(s,p)\\
 & \forall s\in sauce: \forall p1\in pasta, \forall p2\in pasta:  pairedwith(s,p1)\land pairedwith(s,p2)\limplies p1=p2. \\
 & \forall s\in sauce: \forall p\in person, \forall c\in price: chose(s,p)\land payed(p,c)\limplies priced(s,c).  
\end{align*}

\section{Problem definition}\label{sec:problem-definition}
The overarching goal of this paper is to generate a sequence of small reasoning steps, each with an interpretable explanation. 
We first introduce the concept of an explanation of a single reasoning step, after which we introduce a cost function as a proxy for the interpretability of a reasoning step, and the cost of a sequence of such steps. 

\subsection{Explanation of reasoning steps}
We assume that a theory $\allconstraints$ and an initial partial interpretation $I_0$ are given and fixed. 

\begin{definition}
We define the \textbf{maximal consequence} of a theory $\allconstraints$ and partial interpretation $I$ (denoted $max(I,T)$) as the precision-maximal partial interpretation $J$ such that  $I \wedge \allconstraints \entails J$. 
\end{definition}

Phrased differently, $max(I,T)$ is the intersection of all models of $T$ that are more precise than $I$; this is also known as the set of \emph{cautious consequences} of $T$ and $I$ and corresponds to ensuring \emph{global consistency} in constraint solving.
Algorithms for computing cautious consequences without explicitly enumerating all models exist, such as for instance the ones implemented in clasp \cite{DBLP:conf/lpnmr/GebserKS09} or \idp \cite{IDP} (in the latter system the task of computing all cautious consequences is called \emph{optimal-propagate} since it performs the strongest propagation possible).

Weaker levels of propagation consistency can be used as well, leading to a potentially smaller maximal consequence interpretation $max_{other-consistency}(I,T)$. 
The rest of this paper assumes we want to construct a sequence that starts at $I_0$ and ends at $max(I_0,\allconstraints)$ for some consistency algorithm, i.e., that can explain all computable consequences of $\allconstraints$ and $I_0$. 
\begin{definition}
A \textbf{sequence of incremental partial interpretations} of a theory $\allconstraints$ with initial partial interpretation $I_0$ is a sequence $\langle I_0, I_1, \ldots, I_n  = max(I_0,\allconstraints)\rangle$ where $\forall i>0, I_{i-1} \leqp I_{i}$ (i.e., the sequence is precision-increasing).
\end{definition} 

The goal of our work is not just to obtain a sequence of incremental partial interpretations, but also for each incremental step $\langle I_{i-1}, I_i \rangle$ we want an explanation $(E_i,S_i)$ that justifies the newly derived information $N_i = I_i \setminus I_{i-1}$. When visualized, such as in Figure~\ref{fig:zebrascreen}, it will show the user precisely which information and constraints were used to derive a new piece of information.

\begin{definition}
 Let $I_{i-1}$ and $I_i$ be partial interpretations such that $I_{i-1}\land \allconstraints \models I_i$.
 We say that $(E_i,S_i,N_i)$ \emph{explains} the derivation of $I_{i}$ from $I_{i-1}$ if the following hold:
\begin{itemize}
    \item $N_i= I_i \setminus I_{i-1}$ (i.e., $N_i$ consists of all newly defined facts), 
	\item $E_i\subseteq I_i$ (i.e., the explaining facts are a subset of what was previously derived),
	\item $S_i \subseteq \allconstraints$ (i.e., a subset of the constraints used), and 
	\item $S_i \cup E_i \entails N_i$ (i.e., all newly derived information indeed follows from this explanation).
\end{itemize}
\end{definition}

The problem of simply checking whether $(E_i,S_i,N_i)$ explains the derivation of $I_{i}$ from $I_{i-1}$ is in co-NP since this problem can be performed by verifying that $S_i \land \lnot N_i$ has no models more precise than $E_i$. It is hence an instance of the negation of a first-order model expansion problem \cite{DBLP:conf/lpar/KolokolovaLMT10}.

Part of our goal of finding easy to interpret explanations is to avoid redundancy. 
That is, we want a non-redundant explanation $(E_i,S_i,N_i)$ where none of the facts in $E_i$ or constraints in $S_i$ can be removed while still explaining the derivation of $I_i$ from $I_{i-1}$; in other word: the explanation must be \textit{subset-minimal}. 
\begin{definition}
 We call $(E_i,S_i,N_i)$ a \emph{non-redundant explanation of  the derivation of $I_i$ from $I_{i-1}$} if it explains this derivation and whenever $E'\subseteq E_i; S'\subseteq S_i$ while $(E',S',N_i)$ also explains this derivation, it must be that $E_i=E', S_i=S'$. 
\end{definition}

\begin{definition} \label{def:nonred}
A \textbf{non-redundant explanation sequence} is a sequence 
\[\langle(I_0,(\emptyset,\emptyset,\emptyset)), (I_1,(E_1,S_1,N_i)), \dots ,(I_n,(E_n,S_n,N_n))\rangle\]
such that $(I_i)_{i\leq n}$ is sequence of incremental partial interpretations and each $(E_i,S_i,N_i)$ explains the derivation of $I_i$ from $I_{i-1}$.
\end{definition} 

\subsection{Interpretability of reasoning steps}
While subset-minimality ensures that an explanation is non-redundant, it does not quantify how \textit{interpretable} a explanation is. 
This quantification is problem-specific and is often designed in a subjective manner. 
For this, we will assume the existence of a cost function $f(E_i,S_i,N_i)$ that quantifies the interpretability of a single explanation. 

In line with the goal of ``simple enough for a person to understand'' and Occam's Razor, we reason that smaller explanations are easier to interpret than explanations that use a larger number of facts or constraints.
In Section~\ref{sec:cost} we provide a size-based cost function for use in our logic grid puzzle tool, though others can be used as well.
Developing generic methods to find a \emph{good} cost function that correctly captures perceived complexity is an orthogonal research topic we are not concerned with in the current paper. 

\subsection{Interpretability of a sequence of reasoning steps}
In its most general form, we would like to optimize the understandability of the entire sequence of explanations. 
While quantifying the interpretability of a single step can be hard, doing so for a sequence of explanations is even harder. For example, is it related to the most difficult step or the average difficulty, and how important is the ordering within the sequence?
As a starting point, we here consider the total cost to be an aggregation of the costs of the individual explanations, e.g. the average or maximum cost.

\begin{definition}
Given a theory $\allconstraints$ and initial partial interpretation $I_0$, the \textbf{explanation-production problem} consist of finding a non-redundant explanation sequence
\[\langle(I_0,(\emptyset,\emptyset,\emptyset)), (I_1,(E_1,S_1,N_1)), \dots ,(I_n,(E_n,S_n,N_n))\rangle\]
such that a predefined aggregate over the sequence $\left(f(E_i,S_i,N_i)\right)_{i\leq n}$ is minimised.
\end{definition} 

Example aggregation operators are $max()$ and $average()$, which each have their peculiarities: the $max()$ aggregation operator will minimize the cost of the most complicated reasoning step, but does not capture whether there is one such step used, or multiple. Likewise, the $average()$ aggregation operator will favour many simple steps, including splitting up trivial steps into many small components if the constraint abstraction allows this.
Even for a fixed aggregation operator, the problem of holistically optimizing a sequence of explanation steps is much harder than optimizing the cost of a single reasoning step, since there are exponentially more sequences.

\section{Nested Explanations}\label{sec:nested-explanation}

Each explanation in the sequence will be non-redundant and hence as small as possible. Yet, in our earlier work we noticed that some explanations were still quite hard to understand, mainly since a clue had to be combined with implicit constraints and a couple of previously derived facts. All these things \textit{together} implied a consequence, and they had to be taken into account at once.
Such steps turned out to be too complicated to be understood easily and thus require being explained in more detail.

\begin{figure}[t!]
    \centering
    \includegraphics[width=\textwidth]{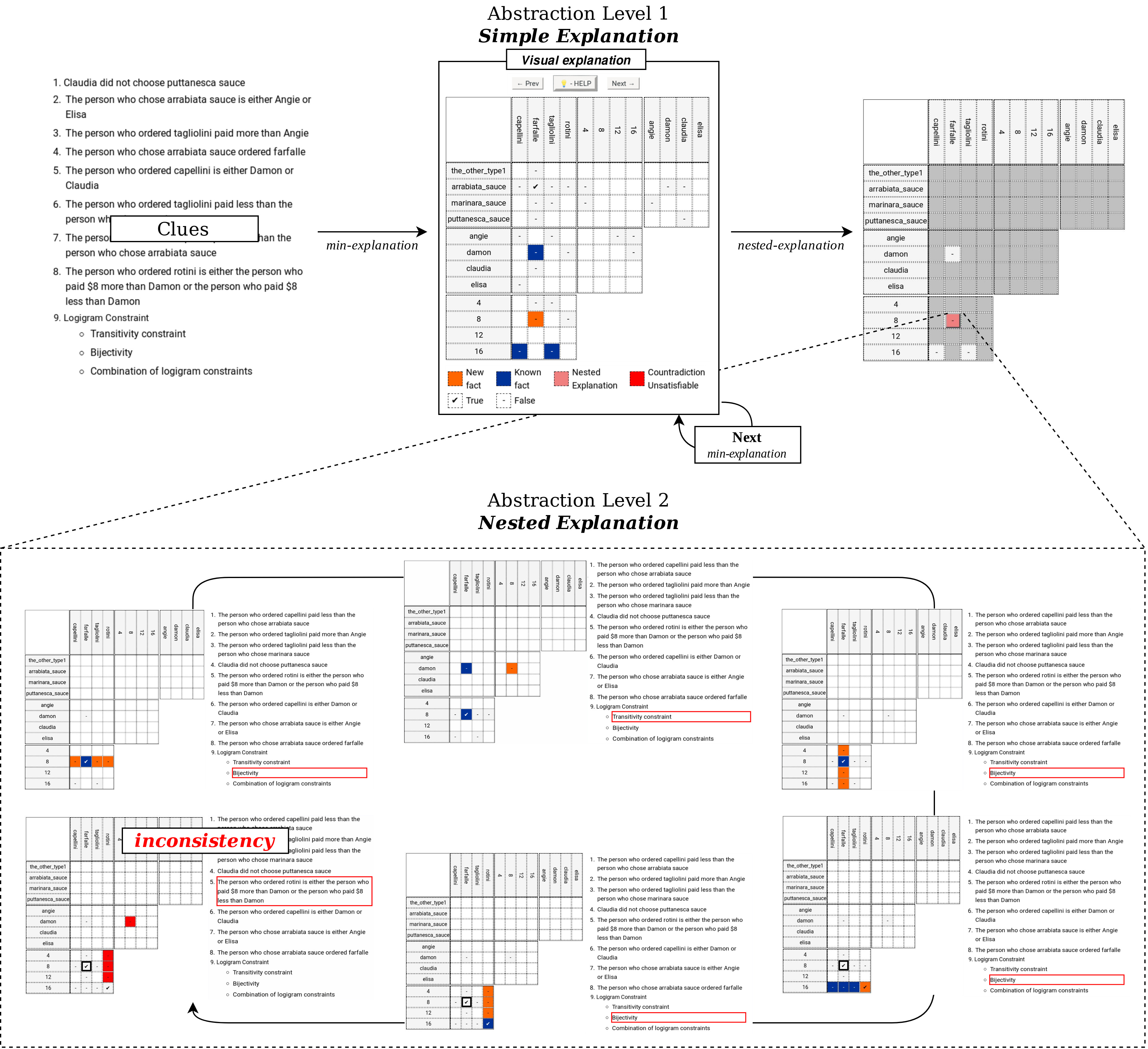}
    \caption{A difficult explanation step, including its nested explanation}\label{fig:pasta_diff}
\end{figure}

An example is depicted at the top in Figure \ref{fig:pasta_diff}.
It uses a disjunctive clue (``The person who ordered Rotini is either the person who paid \$8 more than Damon or the person who paid \$8 less than Damon''), in combination with three previously derived facts to derive that Farfalle does not cost \$8.
This derivation is non-trivial, but can be explained in a step-wise manner using reasoning by contradiction:
\begin{itemize}
    \item If Farfalle did cost \$8, then (since Damon did not eat Farfalle), Damon did not pay \$8;
    \item If Farfalle costs \$8, then it does not cost \$16;
    \item Since Farfalle does not cost \$16 and neither does Capellini or Tagliolini, Rotini must cost \$16;
    \item However, the fact that Rotini costs \$16, while Damon did not pay \$8 is in contradiction with the clue in question;
    \item Hence, Farfalle can not cost \$8.
\end{itemize}
The reasoning step in this Figure~\ref{fig:pasta_diff} is equally straightforward for a computer as the bijectivity reasoning step in Figure~\ref{fig:zebrascreen}. However, understanding the former reasoning step is notably harder for a person.

We hence wish to provide a further, \textit{nested} explanation of such difficult reasoning steps. We believe that an explanation using contradiction is a good tool for this for two reasons: \emph{(i)} it is often used by people when solving puzzles, as well as by mathematicians when proving theorems; and \emph{(ii)} adding the negation of a derived fact such as `Farfalle does not cost \$8', allows us to generate a new sequence of non-redundant explanations up to inconsistency and hence contradiction, hence reusing the techniques from the previous section.
This novel approach allows us to provide a mechanism for \emph{zooming in} into the most difficult explanation step.

\myparagraph{Nested explanation of a reasoning step}

We propose the following principles for what constitutes a meaningful and simple nested explanation, given a non-trivial explanation $(E,S,N)$:
\begin{itemize}
    \item a nested explanation starts from the explaining facts $E$,
          augmented with the counterfactual assumption of a newly derived fact $n \in N$;
    \item at each step, it only uses clues from $S$;
    \item each step is easier to understand (has a strictly lower cost) than the parent explanation with cost $f(E,S,N)$;
    \item from the counterfactual assumption, a contradiction is derived.
\end{itemize}

Note that if an explanation steps derives multiple new facts, e.g. $|N| > 1$, then we can compute a nested explanation for each $n_i \in N$.

More formally, we define the concept of \emph{nested explanation} as follows:

\begin{definition}\label{def:nested-problem}
    The \textbf{nested explanation} problem consists of --- given a non-redundant explanation $(E, S, N)$, and a newly derived fact $n \in N$ --- finding a non-redundant explanation sequence
    \[\langle \ (I_0',(\emptyset,\emptyset,\emptyset)),\ (I_1',(E_1',S_1',N_1')), \dots ,\ (I_n',(E_n',S_n',N_n')) \ \rangle\]
    such that:
    \begin{itemize}
        \item $I_0'$ is the partial interpretation $\{ \neg n_i \wedge E \}$;
        \item $S_i'\subseteq S$ for each $i$;
        \item $f(E_i',S_i',N_i')< f(E, S, N)$ for each $i$;
        \item $I_n'$ is inconsistent; and
        \item a predefined aggregate over the sequence $\left(f(E_i',S_i',N_i')\right)_{i\leq n}$ is minimised.
    \end{itemize}
\end{definition}

We can hence augment each explanation $(E,S,N)$ with a set of nested explanations if they exist. We next discuss algorithms for computing explanations and nested explanations.

\section{Explanation-producing search}\label{sec:expl-gen-prod}
In this section, we tackle the goal of searching for a non-redundant explanation sequence that is as simple as possible to understand.

Ideally, we could generate all explanations of each fact in $max(I_0,\allconstraints)$, and search for the lowest scoring sequence among those explanations. However, the number of explanations for each fact quickly explodes with the number of constraints, and is hence not feasible to compute. Instead, we will iteratively construct the sequence, by generating candidates for a given partial interpretation and searching for the smallest one among those.

\myparagraph{Sequence construction}
We aim to minimize the cost of the explanations of the sequence, measured with an aggregate over individual explanation costs $f(E_i, S_i, N_i)$ for some aggregate like $max()$ or $average()$. The cost function $f$ could for example be a weighted sum of the cardinalities of $E_i$, $S_i$ and $N_i$; in Section~\ref{sec:cost}, we discuss a concrete cost function for the use case of logic grid puzzles.

Instead of globally optimizing the aggregated sequence cost, we encode the knowledge that we are seeking a sequence of small explanations in our algorithm. Namely, we will greedily and incrementally build the sequence, each time searching for the lowest scoring next explanation, given the current partial interpretation. Such an explanation always exists since the end point of the explanation process $max(I_0,\allconstraints)$ only contains consequences of $I_0$ and \allconstraints.

Algorithm \ref{alg:main} formalizes the greedy construction of the sequence, which determines $I_{end} = max(I_0,\allconstraints)$  through propagation and relies on a \textit{min-explanation$(I,\allconstraints)$} function to find the next cost-minimal explanation.

\begin{algorithm}
  $I_{end} \gets$ propagate$(I_0\land \allconstraints)$\;
  Seq $\gets$ empty sequence\;
  $I \gets I_0$\;
  \While{$I \neq I_{end}$}{
    $(E, S, N) \gets $min-explanation$(I, \allconstraints)$\;
    append $(E, S, N)$ to Seq\;
    $I \gets I \cup N$\;
  }
  \caption{High-level greedy sequence-generating algorithm.}
  \label{alg:main}
\end{algorithm}
\myparagraph{Candidate generation}
The main challenge is finding the lowest scoring explanation, among all reasoning steps that can be applied for a given partial interpretation $I$. We first study how to \textit{enumerate} a set of candidate non-redundant explanations given a set of constraints.

For a set of constraints $\allconstraints$, we can first use propagation to get the set of new facts that can be derived from a given partial interpretation $I$ and the constraints $\allconstraints$.
For each new fact $n$ not in $I$, we wish to find a non-redundant explanation $(E \subseteq I, S \subseteq \allconstraints,\{n\})$ that explains $n$ (and possibly explains more).
Recall from Definition~\ref{def:nonred} that this means that whenever one of the facts in $E$ or constraints in $\allconstraints$ is removed, the result is no longer an explanation.
We now show that this task is equivalent to finding a Minimal Unsat Core (or Minimal Unsat Subset, MUS) of a derived this.
To see this, consider the theory
\[ I\wedge \allconstraints \wedge \lnot n.\]
This theory surely is unsatisfiable since $n$ is a consequence of $I$ and $\allconstraints$.
Furthermore, under the assumption that $I\wedge \allconstraints$ is consistent (if it were not, there would be nothing left to explain),
\emph{any} unsatisfiable subset of this theory contains $\lnot n$.
We then see that each unsatisifiable subset of this theory is of the form $E \wedge S \wedge \lnot n$ where $(E,S,\{n\})$ is a (not necessarily redundant) explanation of the derivation of $\{n\}$ from $I$.
Vice versa, each explanation of $\{n\}$ corresponds to an unsatisifiable subset. Thus, the \emph{minimal} unsatisifiable subsets (MUS) of the above theory are in one-to-one correspondence with the non-redundant explanations of $n$, allowing us to use existing MUS algorithms to search for non-redundant explanations.

We must point out that MUS algorithms typically find \textit{an} unsatisfiable core that is \textit{subset-minimal}, but not \textit{cardinality-minimal}. That is, the unsat core can not be reduced further, but there could be another minimal unsat core whose size is smaller.
That means that if size is taken as a measure of simplicity of explanations, we do not have the guarantee to find the optimal ones. And definitely, when a cost function kicks, optimality is also not guaranteed.

Algorithm~\ref{alg:cand} shows our proposed algorithm. The key part of the algorithm is on line \ref{line:mus} where we find an explanation of a single new fact $n$ by searching for a MUS that includes $\neg n$.
We search for subset-minimal unsat cores to avoid redundancy in the explanations.
Furthermore, once a good explanation $(E,S,N)$ is found, we immediately explain all implicants of $E$ and $S$. In other words: we take $N$ to be subset-maximal.
The reason is that we assume that all derivable facts that use the same part of the theory and the same part of the previously derived knowledge probably require similar types of reasoning and it is thus better to consider them at once.
Thus, we choose to generate candidate explanations at once for all implicants of $(E, S)$ on line~\ref{line:implicants}.
Note that the other implicants $N \setminus \{n\}$ may have simpler explanations that may be found later in the for loop, hence we do not remove them from $J$.

\begin{algorithm}
  \SetKwInOut{Input}{input}\SetKwInOut{Output}{output}
  \SetKwComment{command}{/*}{*/}

  \Input{A partial interpretation $I$ and a set of constraints $\allconstraints$}

  Candidates $\gets \{\}$\;
  $J \gets$ propagate$(I \wedge \allconstraints)$\;
  \For{$n \in J \setminus I$\label{line:for}}{
    \tcp{\small Minimal expl. of each new fact:}
    $X \gets MUS(\neg n \wedge I \wedge \allconstraints)$ \label{line:mus}\;
    $E \gets I \cap X$\tcp*{\small facts used}
    $S \gets \allconstraints \cap X$\tcp*{\small constraints used}
    $N \gets$ propagate$(E \wedge S)$\tcp*{\small all implied facts}\label{line:implicants}
    add $(E, S, N)$ to Candidates
  }
  \Return{Candidates}
  \caption{candidate-explanations$(I,\allconstraints)$}

  \label{alg:cand}
\end{algorithm}

We assume the use of a standard MUS algorithm, e.g., one that searches for a satisfying solution and if a failure is encountered, the resulting Unsat Core is shrunk to a minimal one~\cite{marques2010minimal}. While computing a MUS may be computationally demanding, it is far less demanding than enumerating all MUS's (of arbitrary size) as candidates.

\myparagraph{Cost functions and cost-minimal explanations}
We use Algorithm~\ref{alg:cand} to generate candidate explanations, but in general our goal is to find cost-minimal explanations. In the following, we assume that have a cost function $f$ is fixed that returns a score for every possible explanation $(E, S, N)$.

To guide the search to cost-minimal MUS's, we use the observation that typically a small (1 to a few) number of constraints is sufficient to explain the reasoning. A small number of constraints is also preferred in terms of easy to understand explanations, and hence have a lower cost. For this reason, we will  not call \textit{candidate-explanations} with the full set of constraints \allconstraints, but we will iteratively grow the number of constraints used.

We make one further assumption to ensure that we do not have to search for candidates for all possible subsets of constraints. The assumption is that we have an optimistic estimate $g$ that maps a subset $S$ of $\allconstraints$ to a real number such that  $\forall E, N, S: g(S) \leq f(E, S, N)$. This is for example the case if $f$ is an additive function, such as $f(E, S, N) = f_1(E) + f_2(S) + f_3(N)$ where $g(S) = f_2(S)$ assuming $f_1$ and $f_3$ are always positive.

We can then search for the smallest explanation among the candidates found, by searching among increasingly worse scoring $S$ as shown in Algorithm~\ref{alg:minexpl}. This is the algorithm called by the iterative sequence generation (Algorithm \ref{alg:main}).

\begin{algorithm}
  \SetKwInOut{Input}{input}\SetKwInOut{Output}{output}
  \SetKwComment{command}{/*}{*/}
  \SetKw{Break}{break}

  \Input{A partial interpretation $I$ and a set of constraints $\allconstraints$}
  $\mathit{best}\gets none$\;
  \For{$S \subseteq \allconstraints$ ordered ascending by $g(S)$}{ \label{alg:min:for}
    \If{$best\neq none$ {\bf and} $g(S) > f(\mathit{best})$ }{
      \Break\;}
    cand $\gets$ best explanation from candidate-explanations$(I, S)$\;
    \If{$\mathit{best} = none$ {\bf or} $f(\mathit{best}) > f(cand)$}{$ \mathit{best}\gets cand$\; \label{alg:min:gets}}
  }
  \Return{$\mathit{best}$}
  \caption{min-explanation$(I,\allconstraints)$}
  \label{alg:minexpl}
\end{algorithm}

Every time \textit{min-explanation$(I,\allconstraints)$} is called with an updated partial interpretation $I$ the explanations should be regenerated. The reason is that for some derivable facts $n$, there may now be a much easier and cost-effective explanation of that fact.
There is one benefit in caching the \textit{Candidates} across the different iterations, and that is that in a subsequent call, the cost of the most cost-effective explanation that is still applicable can be used as a lower bound to start from.
Furthermore, in practice, we cache all candidates and when we (re)compute a MUS for a fact $n$, we only store it if it is more cost-effective than the best one we have previously found for that fact, across the different iterations.

\subsection{Searching nested explanations}

We extend Algorithm \ref{alg:main} to generate new candidate explanations with support for nested explanations as introduced in Section~\ref{sec:nested-explanation}.
Fundamentally, the generated candidate explanations are further decomposed in a nested explanation sequence provided that the sequence is easier than the candidate explanation according to the defined cost function $f$.
We assess the possibility for a nested explanation for every newly derived fact in the candidate explanation.
Similar to Algorithm~\ref{alg:main}, Algorithm~\ref{alg:nested-main} exploits the \textit{min-explanation} function to generate the candidate nested explanations.
The only difference is that after computing each explanation step, also a call to \textit{nested-explanations} (which is found in Algorithm~\ref{alg:nested-explanation}) is done to generate a nested sequence.

The computation of a nested explanation as described in Algorithm~\ref{alg:nested-explanation} also reuses \textit{min-explanation}; but, the main differences with the high level explanation generating algorithm come from the fact that the search space for next easiest explanation-step is bounded by \emph{(i)} the input explanation: it can use only constraints (and facts) from the original explanation, and \emph{(ii)} the cost of the parent explanation is an upper bound on the acceptable costs at the nested level.

Given an explanation step $(E, S, N)$ and a newly derived fact $n \in N$ for which we want a more detailed explanation, Algorithm~\ref{alg:nested-explanation} first constructs a partial interpretation $I'$ formed by the facts in $E$ and the negated new fact $n$.
Then, we gradually build the sequence by adding the newly found explanations $(E', S', N')$ as long as the interpretation is consistent and the explanation is easier than explanation step $(E, S, N)$ (this is in line with Definition~\ref{def:nested-problem} and serves to avoid that the nested explanation is simply a single-step explanation that is equally difficult as the original step.

While Algorithm \ref{alg:nested-explanation} tries to find a nested explanation sequence for each derived fact, it will not find one for each fact due to the if-check at Line \ref{ifcheck}. This check is present to avoid that the nested explanation is as difficult as the high-level step it aims to clarify. This check can kick in for two different reasons. The first reason is that the explanation step at the main level is simply too simple to be further broken up in pieces. For instance the explanation of Figure \ref{fig:zebrascreen} is of that kind: it uses a single bijectivity constraint with a single previously derived fact. Breaking this derivation up in strictly smaller parts would thus not be helpful. But this phenomenen can also occur for very difficult steps: sometimes the best nested explanation of a difficult explanation step contains a step that is as difficult as the high-level step itself. In that case, this is a sign that reasoning by contradiction is not simplifying matters in this step and other methods should be explored to further explain those steps.

\begin{algorithm}[ht]
  $I_{end} \gets$ propagate$(I_0\land \allconstraints)$\;
  $Seq \gets$ empty sequence\;
  $I \gets I_0$\;
  \While{$I \neq I_{end}$}{
  $(E, S, N) \gets $min-explanation$(I, \allconstraints)$\;
  {\color{gray}
  $nested \gets$ nested-explanations$(E, S, N)$\;
  append $((E, S, N),nested)$ to $Seq$\;
  }

  $I \gets I \cup N$\;
  }
  \caption{greedy-explain$(I_0, \allconstraints)$}
  \label{alg:nested-main}
\end{algorithm}

\begin{algorithm}[ht]
  \SetKwInOut{Input}{input}\SetKwInOut{Output}{output}
  \SetKwComment{command}{/*}{*/}
  \SetKw{Break}{break}
  $nested\_explanations \gets$ empty set\;
  \For{$n \in N$}{
    $store \gets true$\;
    $nested\_seq \gets$ empty sequence\;
    $I' \gets E \wedge \neg \{n\} $ \;
    \While{consistent($I'$)}{
      $(E', S', N') \gets $min-explanation$(I', S)$\;
      \If{$f(E', S', N') \geq f(E,S,N)$ \label{ifcheck}}{
        $store \gets false$; \Break\;}
      append $(E', S', N')$ to $nested\_seq$\;
      $I' \gets I' \cup N'$\;

    }
    \If{$store$}{\tcp{\small only when all steps simpler than (E,S,N)}
      append $nested\_seq$ to $nested\_explanations$\;}
  }
  \Return $nested\_explanations$
  \caption{nested-explanations(E, S, N)}
  \label{alg:nested-explanation}
\end{algorithm}

\section{Explanations for logic grid puzzles}\label{sec:zebra}
We instantiated the above described algorithm in the context of logic grid puzzles.
In that setting, for $\allconstraints$, we take $T_P$ for some puzzle $P$, as described in Section \ref{sec:prelims}.
There are three types of constraints in $\allconstraints$: transitivity constraints, bijectivity constraints and clues, where the first two follow the same structure in every puzzle and the clues are obtained in a mostly automatic way (see Section \ref{sec:holistic}).
Before defining a cost-function, and the estimation for $g$ used in our implementation, we provide some observations that drove our design decision.

\paragraph{Observation 1: Propagations from a single implicit constraint are very easy to understand}\label{obs:1:implicitconstraints} Contrary to the clues, the implicit constraints (transitivity/bijectivity) are very limited in form and propagations over them follow well-specified patterns.
For instance in the case of bijectivity, a typical pattern that occurs is that when $X-1$ out of $X$ possible values for a given function have been derived not to be possible, it is propagated that the last value should be true; this is visualized for instance in Figure \ref{fig:zebrascreen}.
Hence, in our implementation, we ensure that they are always performed first. Stated differently, $g$ and $f$ are designed in such a way that $g(S_1)\geq f(E,S_2,N)$  whenever $S_2$ consists of only one implicit constraint and $S_1$ does not.

\paragraph{Observation 2: Clues rarely propagate by themselves}\label{obs:2:cluespropagate}
We observed that the automatically obtained logic representation of clues usually has quite weak (unit) propagation strength in isolation.
This is not a property of the clues, but rather of the final obtained translation. As an example, consider the following sentence:
``The person who ordered capellini is either Damon or Claudia''. From this, a human reasoner might conclude that Angie did not order capellini.
However, the (automatically) obtained logical representation is
\[\exists p\in \mathit{person}: \mathit{ordered}(p, \mathit{capellini})\land (p =  \mathit{Damon}\lor p =  \mathit{Claudia}).\]
This logic sentence only entails that Angie did not order capellini \emph{in conjunction with the bijectivity constraint on $ \mathit{ordered}$}.
In the natural language sentence, this bijectivity is implicit by the use of the article ``The'' which entails that there is a unique person who ordered capellini.

We observed that there is rarely any propagation from sole clues, and that only few implicit constraints are active together with a clue at any time.
Because of this last observation, in our implementation for logic grid puzzles we decided not to consider all subsets of implicit constraints in combination with a clue as candidate sets $S$ in Line \ref{alg:min:for} in Algorithm \ref{alg:minexpl}. Instead, we combine each clue with the entire set of all implicit constraints, subsequently counting on the non-redundance of the explanation (the subset-minimality of the core) to eliminate most of the implicit constraints since they are not used anyway.

\paragraph{Observation 3: Clues are typically used independently from other clues}\label{obs:3:cluesusedindependently}
A next observation is that in all the puzzles we encountered, human reasoners never needed to combine two clues in order to derive new information and that when such propagations are possible, they are quite hard to explain, and can be split up into derivations containing only single clues.
The latter is of course not guaranteed, since one can artificially devise disjunctive clues that do not allow propagation by themselves.
Our algorithms are built to handle this case as well, but it turned out to be not necessary in practice: in the puzzles we tested, we never encountered an explanation step that combined multiple clues.

\paragraph{Observation 4: Previously derived facts are easier to use than clues or implicit constraints}\label{observation4}
Our final observation that drove the design of the cost functions is that using previously derived facts is often easier than using an extra clue or implicit constraint. This might be due to the fact that previously derived facts are of a very simple nature while, even implicit constraints contain quantification and are thus harder to grasp. An additional reason for this perceived simplicity is that the derived facts are visualized in the grid.

\paragraph{A cost function}
With these four observations in mind, we devised $f$ and $g$ as follows (where $nc(S)$ denotes the number of clues in constraint set $S$): \label{sec:cost}
\begin{align*} & f(E,S,N) = basecost(S) + |E| + 5\cdot|S|                     \\
     & g(S) = basecost(S) = \left\{\begin{array}{ll}
        0              & \text{if $|S|=1$ and $nc(S) = 0$} \\
        100            & \text{if $|S|>1$ and $nc(S)=0$}   \\
        100\cdot nc(S) & \text{otherwise}
    \end{array}\right.
\end{align*}

The number $100$ is taken here to be larger than any reasonable explanation size.
The effect of this,  is that we can generate our subsets $S$ in Line \ref{alg:min:for}
of Algorithm \ref{alg:minexpl} in the following order:
\begin{itemize}
    \item First all $S$ containing exactly one implicit constraint.
    \item Next, all $S$ containing exactly all implicit constraints and (optionally) exactly one clue.
    \item Finally, all clue pairs, triples etc. though in practice this is never reached.
\end{itemize}
Summarized, our instantiation for logic grid puzzles differs from the generic methods developed in the previous section in that it uses a domain-specific optimization function $f$ and does not considering all $S$ in Line \ref{alg:min:for}, but only promising candidates based on our observations.

For the complete non-redundant explanation sequence our tool produces on the running example using these scoring functions, we refer to \url{http://bartbog.github.io/zebra/pasta}. An example of the hardest derivation we encountered (with cost 108), as well as its nested explanation, is depicted in Figure \ref{fig:pasta_diff}. It uses several bijectivity constraints for uniqueness of persons, but also for reasoning on the relation between costs and types of pasta, in combination with a clue and three assumptions.

\section{Logic grid puzzles: From natural language clues to typed first-order logic}\label{sec:holistic}
The demo system we developed, called \ourtool, is named after Einstein's zebra puzzle, which is an integrated solution for solving logic grid puzzles, and for explaining, \textit{in a human-understandable way}, how the solution can be obtained from the clues. 
The input to \ourtool is a plain English language representation of the clues and the names of the \textit{entities} that make up the puzzle, e.g  ``Angie'' , ``arrabiata''. 

In typical logic grid puzzles, the entity names are present in the grid that is supplied with the puzzle. For some puzzles, not all entities are named or required to know in advance; a prototypical example is Einstein's Zebra puzzle, which ends with the question ``Who owns the zebra?'', while the clues do not name the zebra entity and the puzzle can be solved without knowledge of the fact there is a zebra in the first place.

The complete specification undergoes the following steps, starting from the input:

\begin{enumerate}
	\item[\bf A]\label{pipeline:stepA:POS} \textbf{Part-Of-Speech tagging}: A Part-Of-Speech (POS) tag is associated with each word using an out-of-the-box POS tagger \cite{DBLP:journals/coling/MarcusSM94}.
	\item[\bf B]\label{pipeline:stepB:chukingLexicon} \textbf{Chunking and lexicon building}: A problem-specific lexicon is constructed.
	\item[\bf C]\label{pipeline:stepC:ChunkedToLogic} \textbf{From chunked sentences to logic}: Using a custom grammar and semantics a logical representation of the clues is constructed
	\item[\bf D]\label{pipeline:stepD:logicToIDP} \textbf{From logic to a complete IDP specification}: The logical representation is translated into the IDP language and augmented with logic-grid-specific information.
	\item[\bf E]\label{pipeline:stepE:explanationGeneration} \textbf{Explanation-producing search in \idp}: This is the main contribution of this paper, as explained in Section~\ref{sec:expl-gen-prod}.
	\item[\bf F]\label{pipeline:stepF:visualisation} \textbf{Visualisation}: The step-by-step explanation is visualised. 
\end{enumerate}

The first three of these steps are related to Natural Language Processing (NLP) and are detailed in section \ref{sec:pipeline:nlp} hereafter. 
Next, we explain in section \ref{sec:pipeline:IDPtoExplanations} how the IDP specification formed in step \ref{pipeline:stepD:logicToIDP} is used to generate the explanations and visualisations in steps \ref{pipeline:stepE:explanationGeneration} and \ref{pipeline:stepF:visualisation} respectively.
An online demo of our system can be found on \url{http://bartbog.github.io/zebra}, containing examples of all the steps (bottom of demo page). 

\subsection{Natural Language Processing}\label{sec:pipeline:nlp}

\subsubsection*{Step A. POS tagging} The standard procedure in Natural Language Processing is to start by tagging each word with its estimated Part-Of-Speech tag (POS tag).
We use the standard English Penn treebank Pos tagset \cite{marcus1993building} together with NLTK's built-in perceptron tagger \footnote{\url{http://www.nltk.org/}} as POS tagger. 
It uses a statistical inference mechanism trained on a standard training set from the Wall Street Journal. 
Since any POS-tagger can make mistakes, we make sure that all of the puzzle's entities are tagged as \textit{noun}.

\subsubsection*{Step B. Chunking and lexicon building} From a natural language processing point of view, the hardest part is step B: automatically deriving the lexicon and building the grammar.
The lexicon assigns a role to different sets of words, while the grammar is a set of rules describing how words can be combined into sentences. 
The goal of this second step is to group the POS-tagged words of the clues into \textit{chunks} that are tagged with lexical categories of which 3 are puzzle-specific: proper nouns for the individual entities that are central to the puzzle, other nouns to groups of entities (like \textit{pasta, sauce}) and transitive verbs that link two entities to each other (e.g Claudia did not \textit{choose} puttanesca sauce). 
The other categories are determiner, number, preposition, auxiliary verb etc.  and contain a built-in list of possible members.  We refer to \cite{msc/Claes17} for full details on the categories.

This process of grouping words is referred to as \textit{chunking}. 
We use NLTK and a custom set of regular expressions for chunking the proper nouns and different types of transitive verbs.
The result is a lexicon where each word or set of words (chunk) is assigned a role based on the POS tags.  
On top of these roles, we defined a puzzle-independent grammar in the Blackburn and Bos framework \cite{Blackburn2005,Blackburn2006}.
The grammar itself was created based on 10 example training puzzles, and tested on 10 different puzzles to ensure genericity \cite{msc/Claes17}. 

\subsubsection*{Step C. From chunked sentences to logic} Next in \ref{pipeline:stepC:ChunkedToLogic}, the lexicon, partly problem agnostic and partly puzzle-specific, is fed into a type-aware variant of the semantical framework of Blackburn \& Bos \cite{Blackburn2005,Blackburn2006}, which translates the clues into Discourse Representation Theory \cite{DRT}.
The typed extension allows us to discover the case where different verbs are used as synonyms for the same inherent relation between two types, e.g. `$ate(person, pasta)$' and `$ordered(person, pasta)$'.

In our system, this is a semi-automated method that suggests a lexicon and lets a user modify and approve it, to compensate for possible ``creativity'' of the puzzle designers who tend to insert ambiguous words, or use implicit background knowledge such as using ``in the morning'' when there is only one time slot before 12:00.

\subsection{From logic to visual explanations}\label{sec:pipeline:IDPtoExplanations}
Once the types are built, we generate a complete specification which is used by the reasoner, the IDP system~\cite{IDP}, to solve the problem. 

\subsubsection*{Step D. From logic to a complete IDP specification}
From the types built in \ref{pipeline:stepC:ChunkedToLogic}, we construct the IDP vocabulary containing: all the types and a relation for each transitive verb or preposition. 
For instance, if the clues contain a sentence ``Claudia did not choose puttanesca sauce'', then the vocabulary will contain a binary relation chose(.,.) with the first argument of type \textit{person} and the second argument of type \textit{sauce}.

After the vocabulary, we construct IDP theories: we translate each clue into IDP language, and  we add implicit constraints and present in logic grid puzzles.
The implicit constraints are stemming from the translation of the clues : our translation might generate multiple relations between two types. 
For instance, if there are clues ``The person who ate taglioni paid more than Angie'' and ``The person who ordered farfalle chose the arrabiata sauce'', then the translation will create two relation \textit{ate} and \textit{ordered} between persons and pasta.
However we know that there is only one relation between two types, hence we add a theory containing synonymy axioms; for this case concretely:
\[ \forall \ x \in person \ \forall \ y \in pasta : ate(x, y) \leftrightarrow ordered(x, y) \]
Similarly, if two relations have an inverse signature, they represent inversion functions, for instance \textit{isLikeBy} and \textit{ordered} in the clues ``Taglioni is liked by Elisa'' and ``Damon ordered capellini''. 
In this case we add constraints of the form 
\[ \forall \ x \in person \ \forall \ y \in pasta : liked(y, x) \leftrightarrow ordered(x, y)\]
Next, we refer back to the end of section \ref{sec:prelims} for examples of the bijectivity and transitivity axioms that link the different relations.

The underlying solver, \idp\cite{IDP} uses this formal representation of the clues both to solve the puzzle and to explain the solution. 
We chose the \idp system as an underlying solver since it natively offers different inference methods to be applied on logic theories, including model expansion (searching for solutions), different types of propagation (we used optimal-propagate here to find $max(I,\allconstraints)$), and unsat-core extraction and offers a lua interface to glue these inference steps together seamlessly \cite{IDP}.

\section{Experiments}\label{sec:experiments}
Using logic grid puzzles as a use-case, we validate the feasibility of finding non-redundant explanation sequences and generating nested explanation sequences.
As data, we use puzzles from Puzzle Baron’s Logic Puzzles Volume 3~\cite{logigrammen}.
The first 10 puzzles were used to construct the grammar; the next 10 to test the genericity of the grammar.
Our experiments below are on test puzzles only; we also report results on the \textit{pasta} puzzle, which was sent to us by someone who did not manage to solve it himself.

As constraint solving engine, we use \idp~\cite{IDP} due to the variety of inference methods it supports natively.
The algorithms themselves are written in embedded LUA, which provides an imperative environment inside the otherwise declarative \idp system.
The code was not optimized for efficiency and can at this point not be used in an interactive setting, as it takes between 15 minutes to a few hours to fully explain a logic grid puzzle.
Experiments were run on an Intel(R) Xeon(R) CPU E3-1225 with 4 cores and 32 Gb memory, running linux 4.15.0 and \idp 3.7.1.

\begin{table}[t]
	\centering
	\begin{tabular}{c|ccc|cc|cccccc}
		\textbf{p} & \textbf{$|$types$|$} & \textbf{$|$dom$|$} & \textbf{$|$grid$|$} & \textbf{\# steps} & $\overline{\text{\textbf{cost}}}$ & \textbf{1 bij.} & \textbf{1 trans.} & \textbf{1 clue} & \textbf{1 clue+i.}  & \textbf{mult i.} & \textbf{mult c.} \\\hline
		1 &       4 &            5 &         150 &    112 &      27.06 &  31.25\% &  50.00\% &    0.89\% &        17.85\%   &  0\%  &  0\%  \\
		2 &       4 &            5 &         150 &    119 &      27.77 &  23.53\% &  57.14\% &    1.68\% &        17.64\%   &  0\%  &  0\%  \\
		3 &       4 &            5 &         150 &    110 &      23.93 &  32.73\% &  51.82\% &    0\% &        15.46\%   &  0\%  &  0\%  \\
		4 &       4 &            5 &         150 &    115 &      24.54 &  27.83\% &  55.65\% &    2.61\% &        13.92\%   &  0\%  &  0\%  \\
		5 &       4 &            5 &         150 &    122 &      24.97 &  24.59\% &  59.02\% &    0.82\% &        15.58\%   &  0\%  &  0\%  \\
		6 &       4 &            5 &         150 &    115 &      22.58 &  26.96\% &  58.26\% &    2.61\% &        12.18\%   &  0\%  &  0\%  \\
		7 &       4 &            5 &         150 &    110 &      26.79 &  35.45\% &  46.36\% &    0.91\% &        17.27\%   &  0\%  &  0\%  \\
		8 &       4 &            5 &         150 &    118 &      26.81 &  33.90\% &  47.46\% &    3.39\% &        15.25\%   &  0\%  &  0\%  \\
		9 &       4 &            5 &         150 &    114 &      24.75 &  28.95\% &  54.39\% &    3.51\% &        13.16\%   &  0\%  &  0\%  \\
		p &       4 &            4 &          96 &     83 &      34.45 &  33.73\% &  40.96\% &    1.20\% &        21.69\%   &  2.41\%  &  0\%
	\end{tabular}
	\caption{Properties of the puzzles, explanation sequences and constraints used in the explanations.}
	\label{table:composition}
\end{table}

\begin{table}[t]
	\centering
	\begin{tabular}{l|c|cccc|ccccc}
		& \multicolumn{5}{c|}{\bf average nr. of facts used} & \multicolumn{5}{c}{\bf \% of explanations with a clue that use \# facts}                                                                                \\
		p & all &  Bij. & Trans. &  Clue & multi i. & 0 facts & 1 facts & 2 facts & 3 facts & $>$3 facts \\\hline
		1 &  1.84 &  2.37 &   2.00 &  0.52 &        - &  66.67\% &  28.57\% &   0\% &   0\% &    4.76\% \\
		2 &  1.85 &  2.50 &   2.00 &  0.61 &        - &  47.83\% &  47.83\% &   0\% &   4.35\% &    0\% \\
		3 &  1.84 &  2.33 &   2.00 &  0.24 &        - &  82.35\% &  11.76\% &   5.88\% &   0\% &    0\% \\
		4 &  1.89 &  2.50 &   2.00 &  0.47 &        - &  68.42\% &  15.79\% &  15.79\% &   0\% &    0\% \\
		5 &  1.85 &  2.50 &   2.00 &  0.35 &        - &  65.00\% &  35.00\% &   0\% &   0\% &    0\% \\
		6 &  1.84 &  2.35 &   2.00 &  0.29 &        - &  76.47\% &  17.65\% &   5.88\% &   0\% &    0\% \\
		7 &  1.88 &  2.31 &   2.00 &  0.75 &        - &  55.00\% &  25.00\% &  10.00\% &  10.00\% &    0\% \\
		8 &  1.86 &  2.58 &   2.00 &  0.18 &        - &  81.82\% &  18.18\% &   0\% &   0\% &    0\% \\
		9 &  1.85 &  2.45 &   2.00 &  0.32 &        - &  78.95\% &  15.79\% &   0\% &   5.26\% &    0\% \\
		p &  1.73 &  2.07 &   2.00 &  0.53 &     4.00 &  68.42\% &  21.05\% &   0\% &  10.53\% &    0\% \\
	\end{tabular}
	\caption{Statistics on number of previously derived facts $|E|$ used in the explanation steps.	}
	\label{table:sequence_level}
\end{table}

\myparagraph{Sequence composition}
We first investigate the properties of the puzzles and the composition of the resulting sequence explanations. The results are shown in Table~\ref{table:composition}. The first column is the puzzle identifier, where the puzzle identified as p is the pasta puzzle, our running example. 
The next 3 columns show properties of the puzzle:
$|type|$ is the number of types (e.g. person, sauce) while $|dom|$ is the number of entities of each type and $|grid|$ is the number of cells in the grid, i.e.\ the number of literals in the maximal consequence interpretation $I_n=max(\emptyset, \allconstraints)$.
Coincidentally, almost all the puzzles have 4 types with domain size 5,  hence 150 cells, except for the pasta puzzle which has a domain size of 4, thus 96 cells.

Columns 5 and 6 show the amount of steps ($\# steps$) in the explanation sequences found, and $\overline{cost}$ is the average cost of an explanation step in the puzzle. The number of inference steps is around 110-120 for all but the pasta puzzle, which is related to the grid size.

The rest of the columns investigate the proportion of inference steps in the explanation sequence, e.g. the trivial steps using just one bijection constraint (\textbf{1 bij.}), one transitivity constraint (\textbf{1 trans.}) or one clue (\textbf{1 clue}) and no other constraints; and the more complex inference steps using one clue and some implicit (bijectivity or transitivity) constraint (\textbf{1 clue+i}), multiple implicit constraints (\textbf{mult i.}). 
We can observe (see \ref{obs:1:implicitconstraints}) around 30\% of steps are simple bijectivity steps (e.g. completing a row or column in one relation), around 50\% are transitivity steps (except for the pasta puzzle) and up to 3 \% use just a single clue (see \ref{obs:2:cluespropagate}). 
The majority of the steps involving clues use a more complex combination of a clue with other constraints. 
We see that while our method can combine multiple clues, the explanations generated never require combining multiple clues in one inference step (\textbf{mult c.} always 0, see \ref{obs:3:cluesusedindependently}), that is, the method always find simpler steps involving just one clue. 
Also notably, the puzzles from the booklet never require combining implicit constraints, while the harder pasta puzzle does. 
In general, less then $1/5$th of the explanations actually need to use a clue or a combination of a clue and implicit constraints.
Note that however, for puzzle 3, it is not possible to find new facts based on clue information only, it has to be combined with 1 or multiple constraints.

\begin{figure}[t!]
		\centering
		\begin{subfigure}{.5\textwidth}
				\centering
				\includegraphics[width=0.9\linewidth]{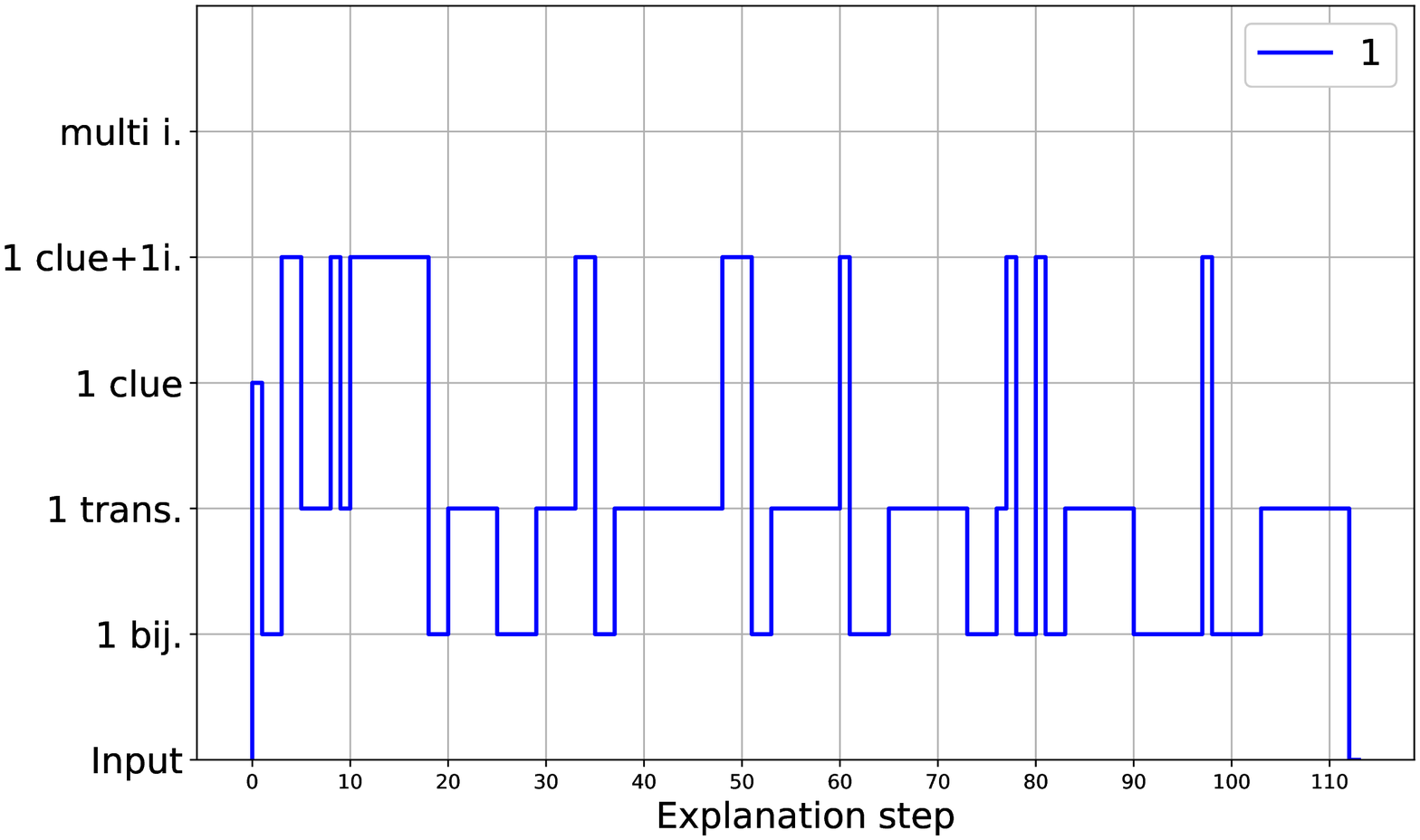}
				\caption{Puzzle 1.}
				\label{fig:composition_puzzle:p1}
		\end{subfigure}%
		\begin{subfigure}{.5\textwidth}
				\centering
				\includegraphics[width=0.84\linewidth]{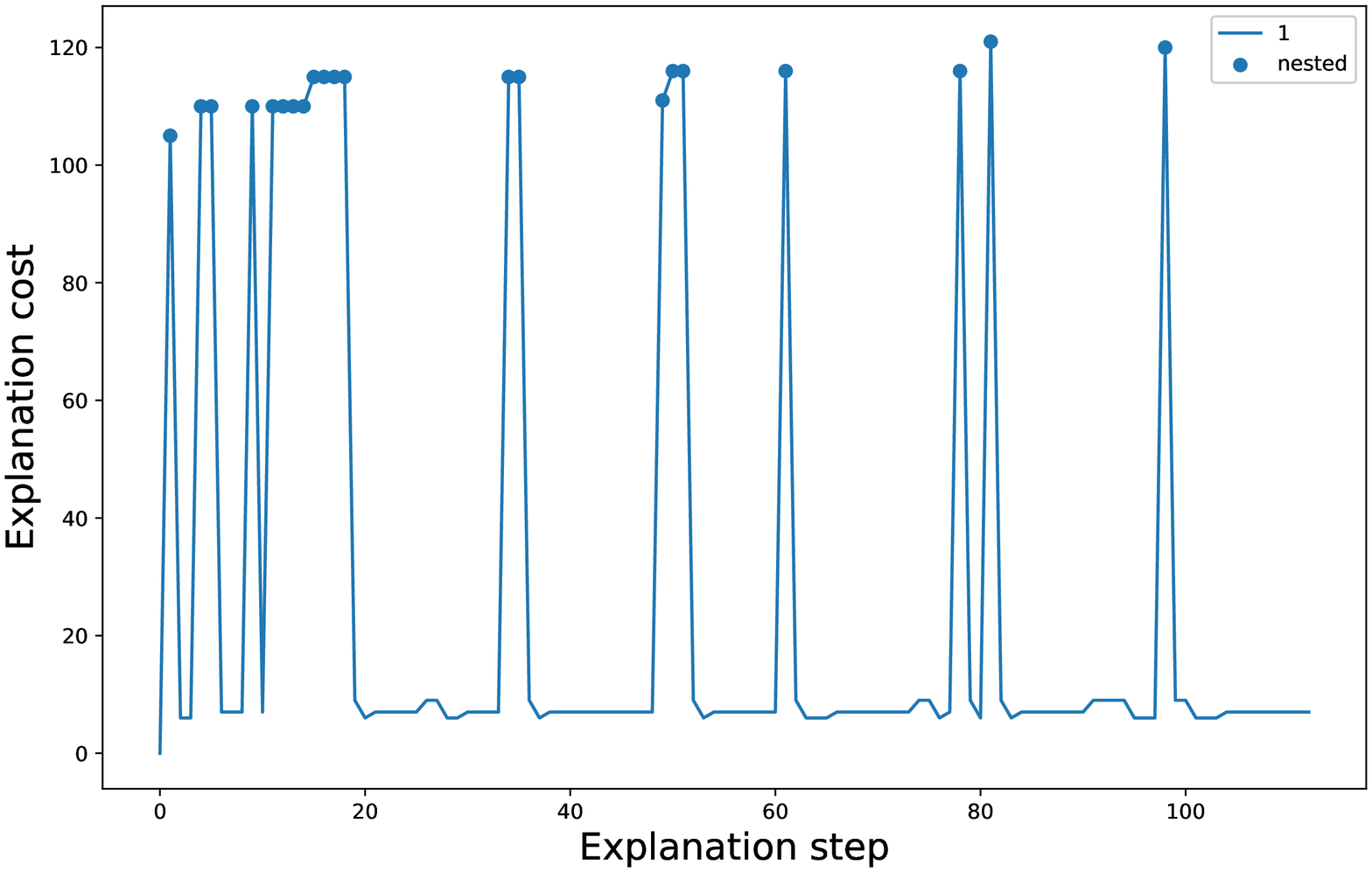}
				\caption{Puzzle 1.}
				\label{fig:cost_puzzle:p1}
		\end{subfigure}
		\begin{subfigure}{.5\textwidth}
				\centering
				\includegraphics[width=0.9\linewidth]{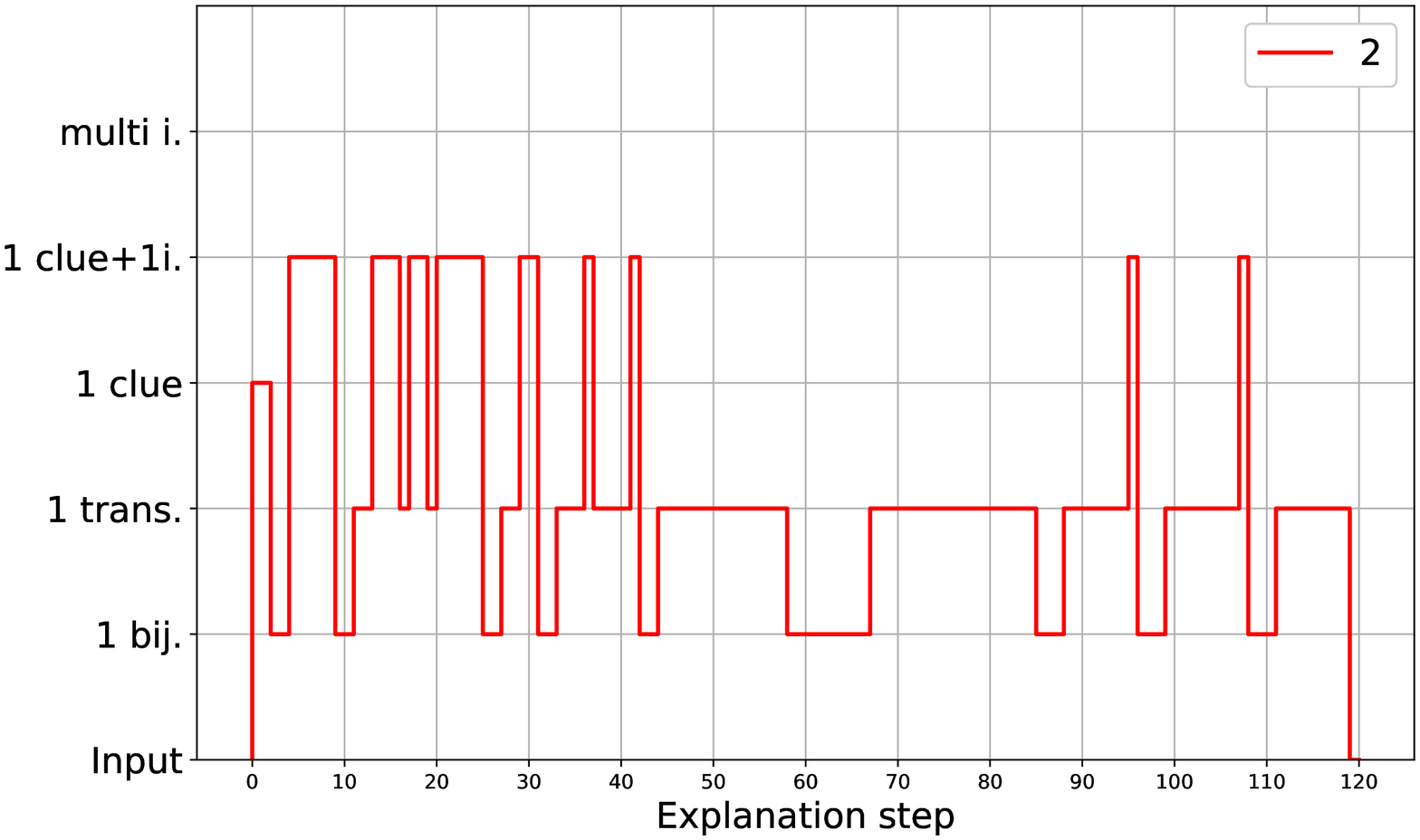}
				\caption{Puzzle 2.}
				\label{fig:composition_puzzle:p2}
		\end{subfigure}%
		\begin{subfigure}{.5\textwidth}
				\centering
				\includegraphics[width=0.84\linewidth]{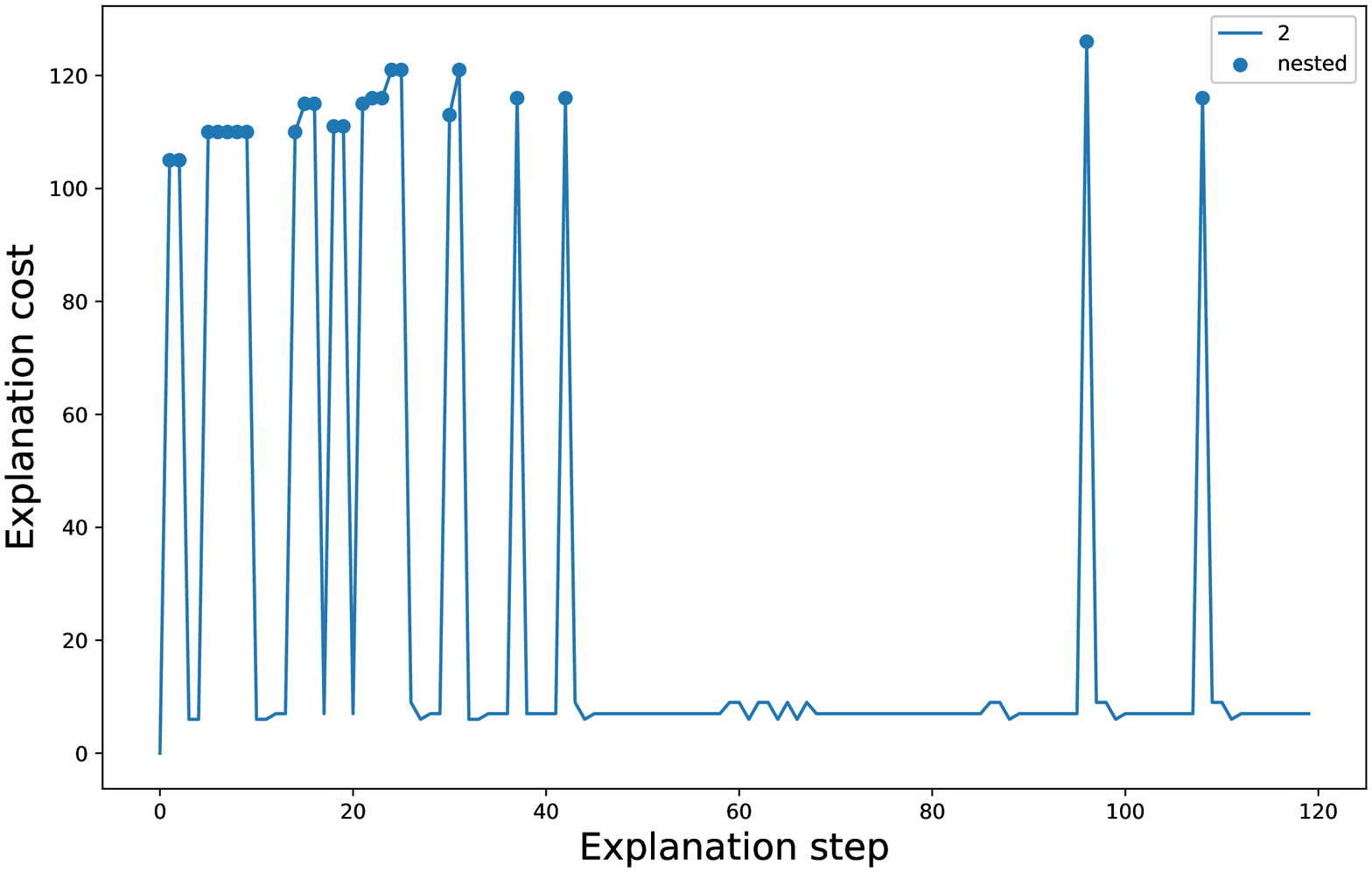}
				\caption{Puzzle 2.}
				\label{fig:cost_puzzle:p2}
		\end{subfigure}
		\begin{subfigure}{.5\textwidth}
				\centering
				\includegraphics[width=0.9\linewidth]{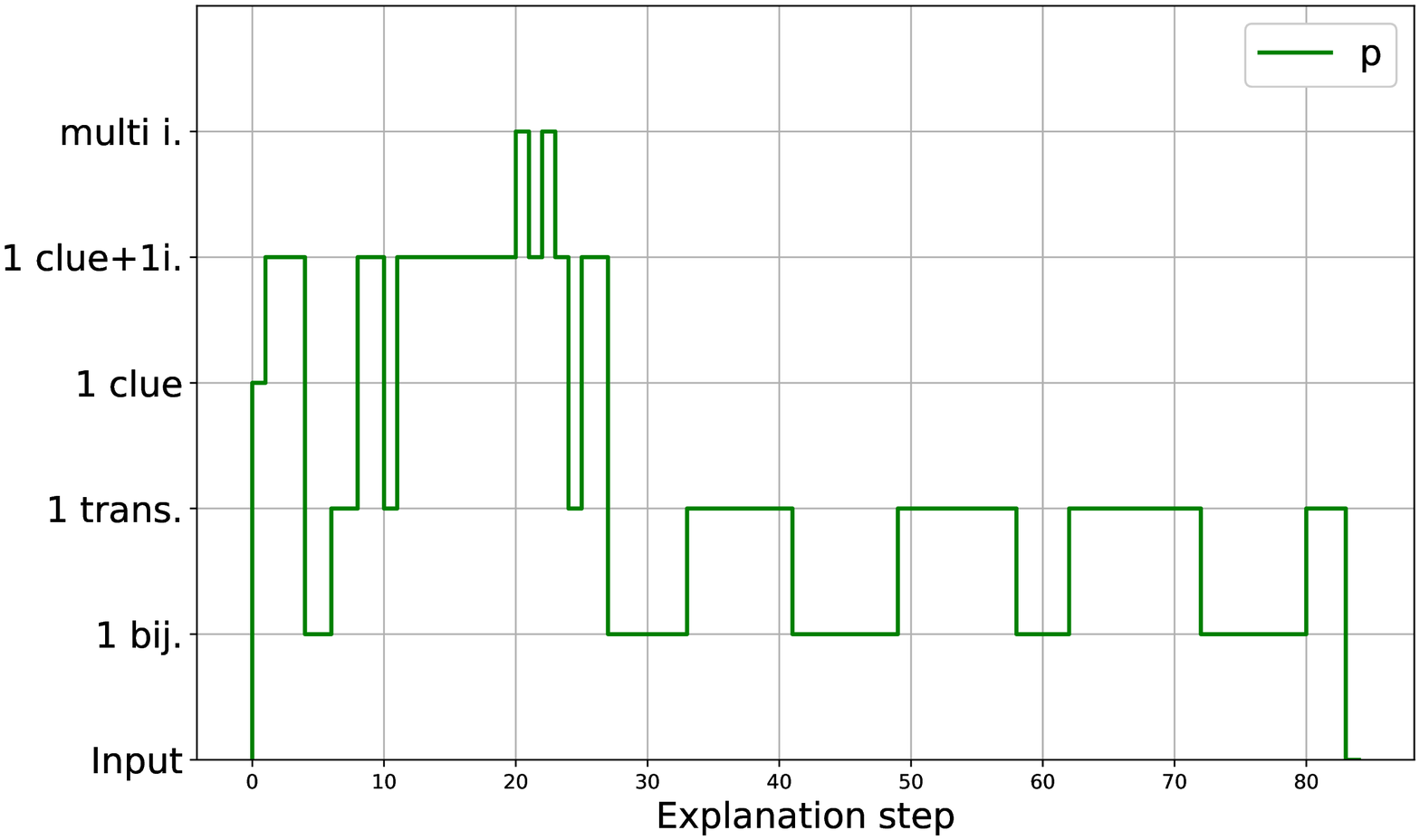}
				\caption{Pasta puzzle.}
				\label{fig:composition_puzzle:pasta}
		\end{subfigure}%
		\begin{subfigure}{.5\textwidth}
				\centering
				\includegraphics[width=0.84\linewidth]{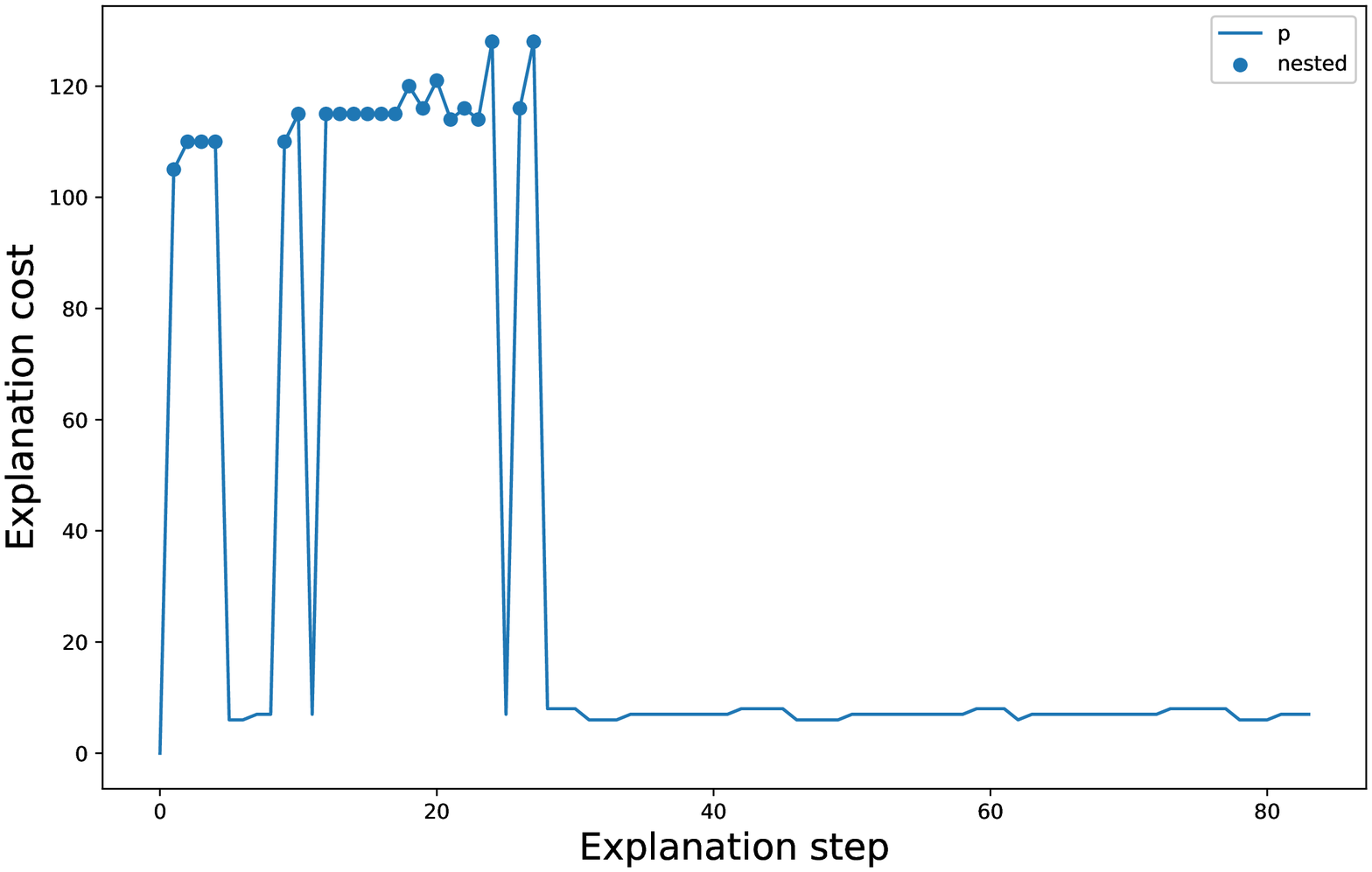}
				\caption{Pasta puzzle.}
				\label{fig:cost_puzzle:pasta}
		\end{subfigure}
		\caption{Side-by-side comparison of puzzle composition (left) and puzzle complexity with nested explanation steps highlighted (right).}
		\label{fig:steps}
\end{figure}

\myparagraph{Sequence progression}
The left side of Figure \ref{fig:steps} shows a visualisation of the type of explanation used in each of the explanation steps for the hardest puzzles 1,2 and p (puzzles with the highest average step cost). 
We can see that typically at the beginning of the sequence, individual clues (3rd line) and some individual bijectivity (1st line) and transitivity (2nd line) constraints are used, i.e., trivial ones.
This is then followed by a series of clues that also involve bijectivity/transitivity constraints, after which a large fraction of the table can be completed with bijectivity/transitivity, followed by a few last clue/implicit constraint combinations and another round of completion.
The exception to this is the pasta puzzle.
We can see that after around 20 steps where mostly clues have been used, twice a combination of implicit logigram constraints must be used to derive a new fact, after which the table can be easily completed with bijectivity/transitivity and twice the use of a clue.

\myparagraph{Explanation size}
Our cost-function is constructed to favour few (if any) clues and constraints in the explanations, and a small number of previously derived facts $|E|$.
Table~\ref{table:sequence_level}, first 5 columns, shows the average number of facts used per type of constraints used in the explanation.
We can observe that the average number of facts used is indeed low, less than two (column 'all').
The breakdown per type of constraint shows that bijectivity typically uses more facts: it either uses three `negative' facts in one row to infer a `positive' fact, as in Figure~\ref{fig:zebrascreen} or it uses one `positive' fact to infer three negative facts.
Note that for such an intuitive constraint, the number of facts used does not matter much. 
Transitivity, by nature, always uses two previously derived facts.
When an explanation involves a clue, few facts are involved on average. 

The rest of the columns take a closer look at the number of facts used when the explanation involves a clue. We can see that our approach successfully finds small explanations: many clues (the trivial ones) use no facts, while some use 1 fact and only occasionally 2 or more facts are needed. 
The puzzles 1, 2, 9 and pasta require the use of 3 facts or more together with a clue corresponding to the puzzles with the highest amounts of clues with implicit constraints. Only puzzle 1 requires clues to be combined with more than 3 facts.
Even though puzzle 9 has a lower cost, the fact it uses more than 3 facts combined with a clue is linked to a complex clue formulation.

Notably the amount of facts for explanations requiring clues, is equal to 0 half of the time, which means that the new information can 
be derived independently of other new facts. 
Part of the clues with 1 fact are linked to the clue involving constraints, namely clues combined with bijectivity as they can derive  3 new facts from only 1 fact.
Altogether clues with 0 or 1 facts form more than 80 \% of the explanations using clues with facts. 

\myparagraph{Nested explanations}

We next investigate the explanation cost of the different steps in an explanation sequence, and which ones we can find a nested explanation for. 
Figure \ref{fig:steps}, right side, 
shows for a few puzzles the explanation cost of each explanation step; for each step it is also indicated whether or not a nested explanation can be found (those where one was found are indicated by a dot in the figure). 
The first observation we can draw from the side-by-side figures, is peaks with nested explanations (on the right) overcome with almost all peaks on the left. Simply put, for each of the non-trivial explanations, we are most often able to find a nested explanation that can provide a more detailed explanation using contradiction. 
Secondly, we observe that the highest cost steps involve more than one constraint: either a clue and some implicit constraints, or a combination of implicit constraints (only in the pasta puzzle). 
The harder pasta puzzle also has a higher average cost, as was already reported in Table~\ref{table:composition}. 

As the explanation sequence progresses, the cost of difficult explanation steps also increases, meaning that we can find easy explanations in the beginning, but we will require more complex reasoning to find new facts that unlock more simpler steps afterwards. 
This is especially visible from the trend line that fits all each puzzle's nested explanation dots on the right side of Figure \ref{fig:steps}. 

\begin{table}[t]
	\centering
	\begin{tabular}{c|c|ccc|c|ccccc}
		& &  \multicolumn{3}{c|}{\textbf{\% steps with nested}} & \multicolumn{1}{c|}{\textbf{average steps}} & \multicolumn{5}{c}{\textbf{Composition of nested explanation }}\\
		\textbf{p} & \textbf{\# steps } & \textbf{of all}     & \textbf{of clue+i.} & \textbf{of m-i}          & \textbf{nested expl.} & \textbf{1 bij.} & \textbf{1 trans.} & \textbf{1 clue} & \textbf{1 clue+i.} & \textbf{mult i.} \\ \hline
		1 &      112 &  14.29\% &  100\% &  - &         2.94 &  36.99\% &   32.88\% &  12.33\% &    17.81\% &    0\% \\
		2 &      119 &  14.29\% &  100\% &  - &         2.65 &  48.28\% &   10.34\% &  20.69\% &    20.69\% &    0\% \\
		3 &      110 &   7.27\% &  100\% &  - &         2.00 &  50.00\% &    0\% &   0\% &    50.00\% &    0\% \\
		4 &      115 &  13.04\% &  100\% &  - &         4.04 &  40.26\% &   29.87\% &  20.78\% &     9.09\% &    0\% \\
		5 &      122 &  13.11\% &  100\% &  - &         2.29 &  47.83\% &    0\% &   8.70\% &    43.48\% &    0\% \\
		6 &      115 &  10.43\% &  100\% &  - &         2.56 &  50.00\% &    8.00\% &  20.00\% &    22.00\% &    0\% \\
		7 &      110 &  15.45\% &  100\% &  - &         3.29 &  37.50\% &   32.69\% &  18.27\% &    11.54\% &    0\% \\
		8 &      118 &   9.32\% &  100\% &  - &         2.36 &  43.75\% &    9.38\% &  25.00\% &    21.88\% &    0\% \\
		9 &      114 &  10.53\% &  100\% &  - &         3.50 &  31.25\% &   28.12\% &  20.31\% &    20.31\% &    0\% \\
		p &       83 &  16.87\% &  100\% &  100\% &         3.07 &  48.44\% &   20.31\% &   7.81\% &    17.19\% &    6.25\% \\
	\end{tabular}
	\caption{Properties of the nested explanations.}
	\label{table:nested_explanation}
\end{table}

We now have a closer look at the composition of these nested explanations in Table~\ref{table:nested_explanation}.
When looking at the percentage of reasoning steps that have a nested explanation, we see that only a fraction of all steps have a nested explanation. 
As the figures already indicated, when looking at the non-trivial steps, we see that for all of those our method is able to generate a nested sequence that explains the reasoning step using contradiction. 

When looking at the number of steps in a nested sequence, we see that the nested explanations are rather small: from 2 to 4 steps.
The subsequent columns in the table show the composition of the nested explanations, in terms of the types of constraints used. 
The majority of those are simple steps involving just a single constraint (a single bijectivity, transitivity or clue). 
Interestingly, even in nested explanations which typically explain a '1 clue+i.' step, the nested explanation also has a step involving a clue and at least one other implicit constraint. 
Detailed inspection showed this was typically the final step, which is a contradiction in the clue as in Figure~\ref{fig:pasta_diff}. 

\begin{figure}[h]
	\centering
	\includegraphics[width=0.6\textwidth]{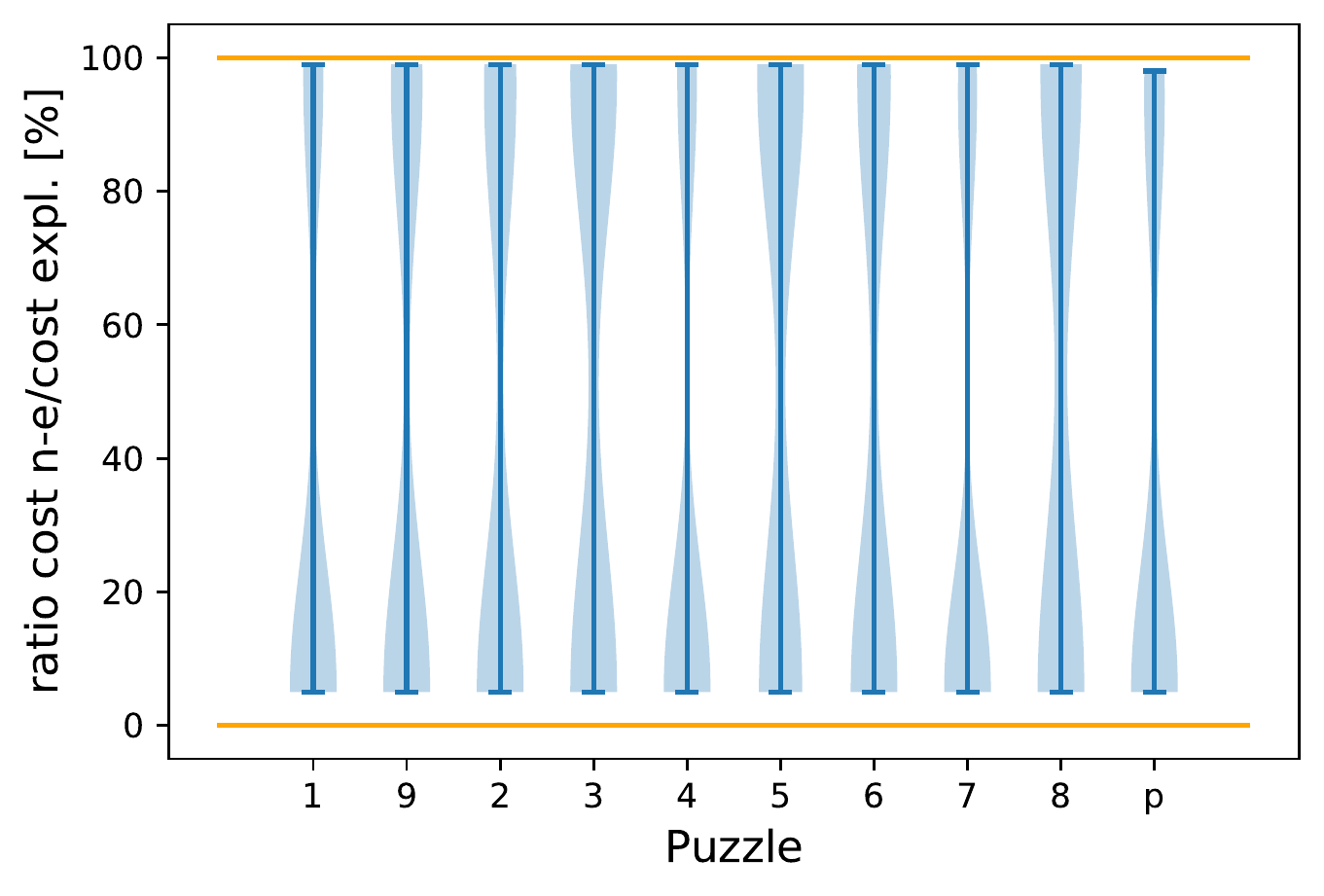}
	\caption{Distribution of the ratio of the cost of a nested explanation sequence step and the cost of original explanation step for all puzzles. 
	} 
	\label{fig:experiments:violin}
\end{figure}

We can further look at the \textit{cost} of the different steps in a nested explanation, which we use as a proxy for difficulty of the steps. 
Figure \ref{fig:experiments:violin} displays a violin plot of the ratio of the cost of a nested step divided by the cost of the original parent step that it is contributing to explain; a wider region indicates a higher density of steps having that ratio. 
Note that by Definition~\ref{def:nested-problem}, a step in a nested explanation can never be as costly or more costly than its parent step (always below the top orange line, though it often comes close).
The figure shows us that there are many steps with a cost of only a fraction of the parent step (near 0-10\% of the original cost), but also some steps closer but still a bit simpler than the parent step. 
Due to the construction of the cost function in Section~\ref{sec:zebra} with each clue contributing a value of '100' and each fact only a value of '1', this means it typically involves fewer previously derived facts, which should make it easier to understand.

\section{Discussion, future work, and conclusions}\label{sec:conclusion}
In this paper, we formally defined the problem of step-wise explanation generation for satisfaction problems, as well as presenting a generic algorithm for solving the problem. We extended the mechanism so that it can be used in a \textit{nested} way, to further explain an explanation.
We developed algorithm in the context of logic grid puzzles where we start from natural language clues and provide a human-friendly explanation in the form of a visualisation. 

When investigating the nested explanation in Figure \ref{fig:zebrascreen:path} as it is produced by the system, one can observe that this explanation does not entirely match an explanation that would be produced by a human reasoner for a couple of reasons: 
\begin{itemize}
 \item In the generation of the nested explanation, as well as in the high-level sequence, we used the greedy algorithm from section \ref{sec:expl-gen-prod}. 
 While at the high level, this yields good results, at the nested level, this results in sometimes propagating facts that are not used afterwards. 
 The very first propagation in the nested explanation is of this kind.
 While this would be easy to fix by postprocessing the generated explanation, we left it in our example to highlight this difference between the nested and non-nested explanation. 
 \item It sometimes happens that the system finds, as a minimal explanation, one in which $X-1$ negative facts are used instead of the corresponding single positive fact. This can be seen in the last step. For human reasoners the positive facts often seem to be easier to grasp. A preference for the system towards these negative facts might be incidentally due to formulation of the clues or it can incidentally happen due to the way the MUS is computed (only subset-minimality is guaranteed there). 
 In general, observations of this kind should be taken into account when devising a cost function. 
 \item A last observation we made (but that is not visible in the current figure) is that sometimes the generated nested explanations seem to be unnecessarily hard. In all cases we encountered where that was the case, the explanation was the same: the set of implicit constraints contains a lot of redundant information: a small number of them would be sufficient to imply all the others. Our cost function, and the subset-minimality of the generated MUS entails that in the explanation of a single step, implicit constraints will never be included if they follow from other included implicit constraints. However, when generating the nested explanations, it would actually be preferred to have those redundant constraints, since they allow breaking up the explanation in simpler parts, e.g., giving a simple step with a single bijectivity, rather than a complex step that uses a combination of multiple implicit constraints.
\end{itemize}
These observations suggest that further research into the question \emph{what constitutes an understandable explanation for humans} is needed. Additional directions to produce easier-to-understand explanations would be optimizing the entire sequence instead of step by step, and learning the cost function based on user traces.

With respect to \emph{efficiency}, the main bottleneck of the current algorithm is the many calls to MUS, which is a hard problem by itself. 
Therefore, in future work we want to investigate unsat-core \emph{optimization} with respect to a cost-function, either by taking inspiration for instance from the MARCO algorithm~\cite{liffiton2013enumerating} but adapting it to prune based on cost-functions instead of subset-minimality, or alternatively by reduction to quantified Boolean formulas or by using techniques often used there~\cite{QBF,DBLP:journals/constraints/IgnatievJM16}.

From a systems point of view, a direction for future work is to make our approach interactive, essentially allowing \ourtool to be called \emph{while} a user is solving the puzzle and to implement in more serious domains such as for instance interactive configuration in which a human and a search engine cooperate to solve some configuration problem and the human can often be interested in understanding \emph{why} the system did certain derivations \cite{DBLP:journals/tplp/HertumDJD17,DBLP:conf/bnaic/CarbonnelleADVD19}. 

Finally, we wish to extend the approach so that it is also applicable to \textit{optimisation problems}. An additional challenge is explaining \textit{search} choices on top of propagation steps as we do in this work. Many open questions remain, but key concepts such as a sequence of simple steps and searching for simple explanations based on explanations of individual facts offer interesting starting points.








\section*{Acknowledgements}
This research received funding from the Flemish Government under the ``Onderzoeksprogramma Artifici\"ele Intelligentie (AI) Vlaanderen'' programme. 
We thank Jens Claes (and his master thesis supervisor Marc Denecker) for the implementation of a typed extension of the Blackburn \& Bos framework as part of his master's thesis, as well as Rocsildes Canoy for his help with the NLP aspect of the information pipeline. 


\bibliography{mybibfile,ref,krrlib}

\end{document}